\documentclass[%
 reprint,
superscriptaddress,
 amsmath,amssymb,
prb,
]{revtex4-2}

\usepackage{graphicx}
\usepackage{dcolumn}
\usepackage{bm}
\usepackage{makecell}
\usepackage{multirow}
\usepackage{mathtools}

\usepackage{comment}
\usepackage{bbold}
\usepackage{xcolor}
\usepackage{bbm}
\usepackage{mathdots}
\usepackage[caption=false]{subfig}
\usepackage{overpic}
\usepackage[T1]{fontenc}
\begin{document}

\preprint{APS/123-QED}

\title{Improved Actions using The Renormalization Group}

\author{Guy Segall}
 \affiliation{Department of Physics, Technion -- Israel Institute of Technology, Haifa, Israel 32000}

\author{Snir Gazit}
\affiliation{The Racah Institute of Physics and The Fritz Haber Research Center for Molecular Dynamics, The Hebrew University of Jerusalem,
Jerusalem 91904, Israel}%

\author{Daniel Podolsky}
\affiliation{Department of Physics, Technion -- Israel Institute of Technology, Haifa, Israel 32000}

\date{\today}

\begin{abstract}
We introduce a numerical method to study critical properties near classical and quantum phase transitions.  Our method applies ideas of the Tensor Renormalization Group to obtain an improved action which is used to extract critical properties by performing Monte Carlo simulations on relatively small system sizes.  We demonstrate this method on the XY model in three dimensions.   Our method may provide a framework with which to efficiently study universal properties in a large class of phase transitions.
\end{abstract}


\maketitle


\section{\label{sec:introduction} Introduction}

One of the most fascinating aspects of continuous phase transitions, both classical and quantum, is their universality.  Seemingly unrelated systems, such as uniaxial ferromagnets, the liquid-gas transition, and binary alloys all share common properties, such as the same critical exponents and amplitude ratios,  belonging to the Ising universality class.  More generally, the universality class of a system is determined by its symmetry and dimensionality and is otherwise independent of microscopic details.

A physical picture of the origin of universality is given by the renormalization group (RG).  The RG involves a sequence of coarse-graining and rescaling transformations, starting from the microscopic model, which allows the extraction of system properties at increasingly longer distances. In this process, almost all of the parameters of the microscopic model flow to zero.  These are the irrelevant operators.  The remaining, relevant, operators are tuned to reach the transition.  Universality is a consequence of the lack of free parameters in the effective model at the critical point.

Critical points are often described by strongly coupled field theories \cite{ssbook}.  This makes standard perturbative approaches unreliable and requires the use of other methods, both analytical and numerical, to obtain critical properties reliably \cite{witczak2014dynamics,chen2014universal,PhysRevB.90.245109,kos2015bootstrapping,ranccon2014higgs,rose2015higgs}.  However, these approaches can struggle in certain applications.  For instance, specific observables can display sizeable corrections to scaling  \cite{wegner1972corrections}, which give large non-universal contributions at short distances.  In Monte Carlo simulations, very large system sizes are required to overcome this, demanding heavy numerical resources to obtain high-accuracy critical exponents and to extract dynamical critical properties near quantum critical points \cite{GPA,GPAA,lucas2017dynamical}. 

In light of this, we pursue a program to systematically reduce the non-universal contributions in numerical studies of phase transitions by using {\em improved actions}.  This idea, first introduced in high energy physics to perform lattice gauge theory calculations \cite{SYMANZIK1983187,alford1995lattice}, exploits the considerable freedom in the choice of lattice models within a given universality class.  In particular, one can add any number of parameters that are irrelevant in the RG sense.  These parameters can be tuned to reduce non-universal contributions, without affecting the universal critical properties. 

In the condensed matter setting, improved actions have previously been used to reduce corrections to scaling.  For instance, in Ref.~\onlinecite{hasenbusch_eliminating_2001}, the coefficient of a single irrelevant perturbation was fixed by scanning over it until the leading correction to scaling was removed.  The net result is an action that well-approximates the continuum limit at the critical point, even at short distances of the order of a few lattice spacings. 

In this paper, we will present a method to generate improved actions without the need to fine-tune model parameters.  The method combines the ideas of the Tensor Renormalization Group (TRG) with Monte Carlo simulations to extract universal critical properties.  We will apply our method to the three-dimensional XY model to obtain critical exponents and correlation functions at criticality.

The TRG is a renormalization group method based on representing the partition function as a tensor network. Traditionally, the TRG involves repeated coarse-graining steps, which are often performed until a fixed point is reached.  Then, properties of interest can be calculated using the coarse-grained tensors. However, this method suffers from biases that originate in the truncation of the tensors after each coarse-graining step, as is necessary to keep the numerical procedure tractable.

Our method involves combining the TRG with Monte Carlo simulations to overcome this bias. Rather than iterating the TRG until a fixed point is found, we perform a small number of TRG iterations and use the resulting coarse-grained tensor partition function as an improved action for Monte Carlo simulations. This allows us to study effectively large systems while simulating small lattices.  In particular, since the improved action is in the same universality class as the original model, Monte Carlo is {\em guaranteed to yield unbiased critical properties}.  Importantly, the improved action benefits from reduced corrections to scaling and other artifacts of performing simulations on a discrete and finite lattice, as we will show.

\section{\label{sec:RGImproved} Renormalization-Group Improved Actions}

In this section, we introduce the idea of Renormalization-Group Improved Actions and provide a brief overview of the method.  For notational simplicity, we present a two-dimensional version on the square lattice.   In Sec.~\ref{sec:methods} we will explain the method in more detail, including how to generalize it to three-dimensional models on the cubic lattice.

\subsection{\label{subsec:TensorRepresentation} Tensor Representation of Partition Functions}

\begin{figure}[b]
    \includegraphics{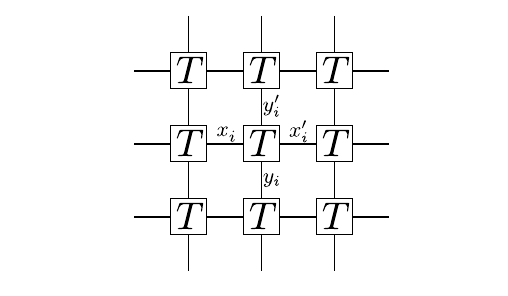}
    \caption{\label{fig:T} The tensor network. Indices are written for the middle tensor, on the site $i$. Prime indices are used for bonds above or to the right of the tensor.  Note that $x_i^\prime= x_{i+\hat{x}}$ and $y_i^\prime= y_{i+\hat{y}}$.}
\end{figure}

Our starting point is a partition function that is represented as a tensor network on the square lattice, of the form, 
\begin{equation}
    Z = \sum_{\{x_i,y_i\}=1}^D
    {\prod_i{T_{x_i x_i^\prime y_i y_i^\prime}}}
    \label{eq:tensorZ}
\end{equation}
where $i$ runs over all lattice sites, $x_i,y_i$ are bond indices of the tensor on site $i$ in the $x$ and $y$ directions, and $D$ is the bond dimension of the indices of the tensor $T$. The notation is such that $x_i^\prime = x_j$ for the next neighboring site $j$ in the $x$ direction, and similarly for the $y$ direction, see Fig.~\ref{fig:T}.

Many quantum and classical statistical mechanics models allow for such a representation.  For concreteness, we will demonstrate this on the XY model.  Consider the soft-spin O$(2)$ model on the cubic lattice, with the Hamiltonian:
\begin{eqnarray}\label{eq:O2model}
    H_{\mathrm{O}(2)} =&& -g\sum_{\langle i,j \rangle}
    {\left (
    \psi_i\psi_j^* + c.c.
    \right )} \nonumber \\
    && + \sum_i{
    \left (
        {\left| \psi_i \right|}^2
        +
        \lambda {\left (
        {\left| \psi_i \right|}^2 - 1
        \right )}^2
    \right )}
\end{eqnarray}
where $\psi_i$ is complex valued.  We perform a high-temperature expansion to rewrite the partition function in the worm representation \cite{Prokofev2001WormAlgorithms}. In the limit where $\lambda\to\infty$ (the hard-spin XY model), this yields:
\begin{equation}\label{eq:Zworm}
    Z = \sum_{CP}{
    \left ( 
        \prod_b{\frac{g^{N_b^1+N_b^2}}{N_b^1!N_b^2!}}
    \right )
    }
\end{equation}
where the sum is over all directed closed paths, the product $\prod_b$ is over all bonds on the lattice, and $N_b^1,N_b^2$ are non-negative integers counting the number of paths going through bond $b$ in each direction.  

Equation \eqref{eq:Zworm} can be cast as a tensor network by defining the tensor:
\begin{equation}
    T_{\{N_b^1,N_b^2\}} = \sqrt{\frac{g^{J_{\text{in}}+J_{\text{out}}}}{\displaystyle \prod_{b\in \mathrm{site}}N_{b}^{1}!N_{b}^{2}!}} \ 
    \delta_{J_{\text{in}},J_{\text{out}}}
    \label{eq:bareT}
\end{equation}
where $b$ runs over all bonds connected to the site of the tensor, $J_{\text{in}}=\sum_b{N_b^1}$ is the total current going into the site, and $J_{\text{out}}=\sum_b{N_b^2}$ is the total current going out of the site. The Kronecker delta function $\delta_{J_{\text{in}},J_{\text{out}}}$ ensures that we sum only over closed paths by requiring conservation of current at each site.  Conservation of current is a direct consequence of the O$(2)$ symmetry to global phase rotations.

This is, in fact, only one of multiple ways of writing the XY model in tensor form.  We choose this representation, based on the worm algorithm \cite{Prokofev2001WormAlgorithms}, because of two advantages.  First, it does not suffer from significant critical slowing down at the phase transition.  Second, it enables keeping exact XY symmetry during the RG transformation -- so long as current conservation is maintained exactly at every step of the RG.

\subsection{Tensor Renormalization Group}

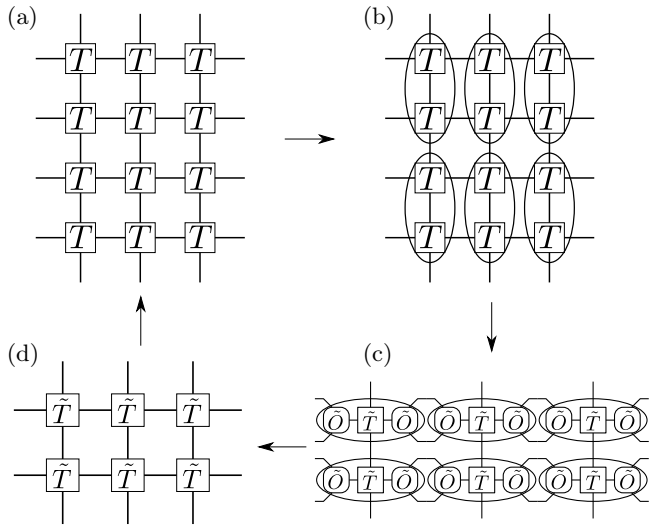
\begin{figure}
    \begin{overpic}[width=.48\textwidth]{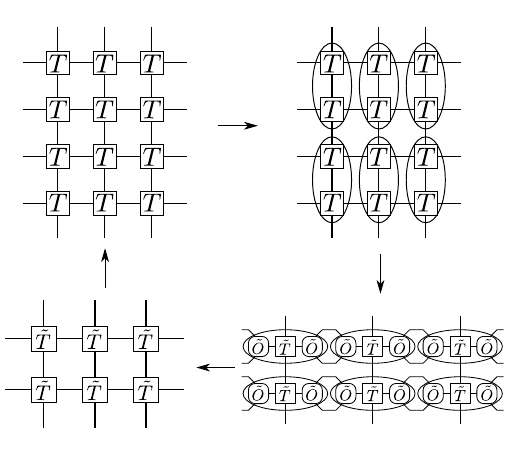}
    \put(0, 86){(a)}
    \put(55, 86){(b)}
    \put(55, 34){(c)}
    \put(0, 34){(d)}
    \end{overpic}
    \caption{\label{fig:TRGcomplete} Schematic description of the Tensor Renormalization Group scheme. (a) The original tensor network. (b) Sites are grouped into supersites, each containing two tensors. (c) Each of the supersites is transformed using the TRG. (d) Contracting the orthogonal tensors, we obtain a new tensor network which forms the basis for the next iteration of the RG.}
\end{figure}

The tensor Renormalization Group (TRG) \cite{Levin2007TensorRenormalizationGroup,Xie2012RGbyHOSVD} is a method for performing real space Renormalization Group transformations on tensor networks. The method is depicted schematically in Fig.~\ref{fig:TRGcomplete}. Starting from the tensor network in Fig.~\ref{fig:TRGcomplete}a, we coarse-grain by contracting tensors in pairs (Fig.~\ref{fig:TRGcomplete}b), then we approximate the resulting tensor using a new core tensor $\tilde{T}$ multiplied by orthogonal tensors $\tilde{O}$, as shown in Fig.~\ref{fig:TRGcomplete}c (we describe this transformation in more detail in Sec.~\ref{sec:methods}). Finally, we contract the orthogonal tensors $\tilde{O}$ to return to a tensor network of the original form, but with new tensors $\tilde{T}$ (Fig.~\ref{fig:TRGcomplete}d). 
This new tensor network forms the basis for the next iteration of the RG. The full TRG procedure is described in more detail in Sec.~\ref{sec:methods}.

In the case of the XY model, current conservation ensures that the XY symmetry is maintained exactly when we coarse-grain and truncate, so that the coarse-grained model belongs to the same universality class as the XY model. 

\subsection{Monte Carlo importance sampling of tensors}

As explained above, we use the tensor network obtained through TRG as an improved action for Monte Carlo simulations. To perform Monte Carlo importance sampling on the tensor network we consider our state as the collection of bond indices $\{x_i,y_i\}$ across the lattice. At each step, one suggests changing the bond indices to new values $\{\tilde{x}_i,\tilde{y}_i\}$. The change is accepted with the Metropolis acceptance ratio, where the weight of each state is given by the tensor network:
\begin{equation}
    P(\{x_i,y_i\} \to \{\tilde{x}_i,\tilde{y}_i\}) =
        \min \left(
        1,
        \frac
        {\prod_i{T_{\tilde{x}_i \tilde{x}_i^\prime \tilde{y}_i \tilde{y}_i^\prime}}}
        {\prod_i{T_{x_i x_i^\prime y_i y_i^\prime}}} 
        \right)
\end{equation}

This requires that the tensors be non-negative, so the tensor network generates non-negative weights. Otherwise, we run into the so-called ``sign problem'' of Monte Carlo.  The standard truncation used in the TRG, using SVD \cite{Xie2012RGbyHOSVD}, does not satisfy this property.  Therefore, we devise an alternative truncation method, the {\em Positive Tensor Truncation} (PTT), that yields non-negative tensors.  Using tensors generated by the PTT we can perform Monte Carlo simulations without a sign problem.

\section{\label{sec:Results} Results}

In this section we show results we obtained in Monte Carlo simulations for the XY model on the cubic lattice using the methods described in Sec.~\ref{sec:methods}. We coarse grained up to five times, and focused on the models after zero, $1$, $3$, and $5$ coarse graining steps. We label these models RG0, RG1, RG3, and RG5, respectively. For each one, we identified the critical coupling $g_c$  and measured observables at and near the critical point. We used lattice sizes between $4$ and $128$.

\subsection{\label{subsec:ScalarObservables} Corrections to Scaling \& Scalar Observables}

We measured the following scalar observables, following \cite{XuWanwan2019HighprecisionMonteCarlostudy}: the superfluid stiffness $\rho_s$, the wrapping probabilites $R_x$, $R_a$, and $R_2$, and the closing time $T_\mathrm{w}$.
The superfluid stiffness can be computed from the worm winding numbers $\mathcal{W}_i$ through, 
    \begin{equation}\label{eq:rhos}
        \rho_s =  \frac{1}{3L}
        \langle
        \mathcal{W}_x^2 + \mathcal{W}_y^2 + \mathcal{W}_z^2
        \rangle
    \end{equation}
    where $\mathcal{W}_i$ is the number of worms winding the entire lattice in direction $i$. The wrapping probabilities are,
    \begin{eqnarray}
        R_x = && \langle \mathcal{R}_x \rangle \label{eq:Rx}\\
        R_a = && 1 - 
        \langle(1-\mathcal{R}_x)(1-\mathcal{R}_y)(1-\mathcal{R}_z)\rangle \label{eq:Ra}\\
        R_2 = && \langle\mathcal{R}_x\mathcal{R}_y(1-\mathcal{R}_z)\rangle
        + \langle\mathcal{R}_x(1-\mathcal{R}_y)\mathcal{R}_z\rangle \nonumber\\
        && + \langle(1-\mathcal{R}_x)\mathcal{R}_y\mathcal{R}_z\rangle \label{eq:R2}
    \end{eqnarray}
    where $\mathcal{R}_i$ is $1$ if $\mathcal{W}_i\neq 0$, and $0$ otherwise. Hence, $R_x$ is the probability for a worm to wrap in the $x$ direction, $R_a$ is the probability that it wraps in any direction, and $R_2$ is the probability that it wraps in exactly two directions.  Finally, the closing time $T_\mathrm{w}$ is a susceptibility-like observable that measures the average time it takes to close a worm:
    \begin{equation}
        T_\mathrm{w} = \langle \tau_\mathrm{w} \rangle
        \label{eq:Tw}
    \end{equation}
    where $\tau_\mathrm{w}$ is the number of worm steps it took to close the worm.

In addition, we measured the derivatives by the coupling $g$ of the superfluid stiffness and the wrapping probabilities. Given an observable $\mathcal{R}$, its derivative $G_R=\frac{d{\langle\mathcal{R}}\rangle}{dg}$ is calculated by
\begin{equation}
    G_R = \langle \mathcal{R}\varepsilon \rangle
     - \langle \mathcal{R}\rangle \langle\varepsilon \rangle
\end{equation}
where $\varepsilon$ is given by
\begin{equation}
    \varepsilon 
    = \sum_i \frac{1}{T_i}\frac{\partial T_i}{\partial g}
\end{equation}
where $T_i$ is the tensor at site $i$: $T_i=T_{x_ix_i^\prime y_i y_i^\prime z_i z_i^\prime}$. The derivative $\frac{\partial T_i}{\partial g}$ is calculated numerically.

\subsubsection{\label{subsubsec:CriticalCoupling} Critical Coupling}

First, we obtained the critical coupling with high precision for every coarse graining step, using the dimensionless observables $\rho_s L,R_x,R_a,R_2$, by fitting each of them to the form:
\begin{equation}\label{eq:CorrectionsToScaling}
    R(g, L) = \sum_{n,m=0}^3 a_{nm}(g-g_c)^n L^{\frac{n}{\nu} - \omega_m}
\end{equation}
where $a_{nm}$ are constants, $g_c$ is the critical coupling, $\nu$ is the critical exponent, for which we used the value $\nu=0.67169(7)$ \cite{Hasenbusch2019MonteCarloClock}, and $\omega_m$ is the exponent of the $m$'th correction to scaling included in the fit. We used $n=0,1,2,3$ and $m=0,1,2,3$ with exponents $\omega_0=0$, $\omega_1=0.789(4)$, $\omega_2=2-\eta$, $\omega_3=2.02(1)$ and $\eta=0.03810(8)$ \cite{Hasenbusch2019MonteCarloClock}. In the fit, we used both the data of the dimensionless observables and the data of their derivatives.

\begin{figure}
    \includegraphics[width=0.9\linewidth]{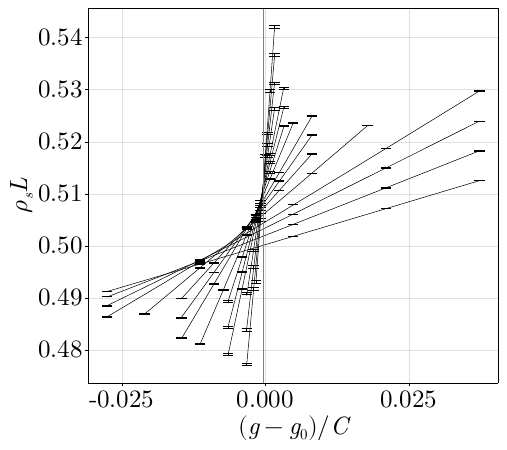}
    \caption{\label{fig:CriticalCouplingFit} $\rho_s L$  as a function of coupling $g$ (shifted and rescaled with arbitrary constants $g_0=0.22717, C=162.1$ for plotting clarity) for various system sizes $L\in[10,12,14,16,20,24,28,32,40,48,56,64,80,96,112,128]$, for the model RG0.  The lines show a joint fit for all lattice sizes to Eq.~\eqref{eq:CorrectionsToScaling}, which is used to extract the critical coupling $g_c$ (grey vertical line).  Slopes increase monotonically with $L$.}
\end{figure}

\begin{figure}
    \includegraphics[width=0.9\linewidth]{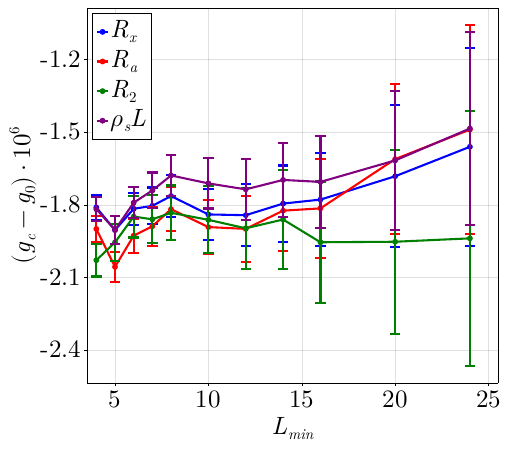}
    \caption{\label{fig:CriticalCouplingVsLminRG0} Critical coupling $g_c$ of RG0 vs minimal lattice size $L_{min}$ used in the fit. The y axis is shifted by $g_0=0.22717$ for plotting clarity.}
\end{figure}

An example of such a fit is shown in Fig.~\ref{fig:CriticalCouplingFit}.%
The fit has a total of $17$ fitting parameters: The critical coupling $g_c$ and the constants $a_{nm}$ for the $16$ pairs ${n,m}$.   Figure~\ref{fig:CriticalCouplingVsLminRG0}
shows the critical coupling $g_c$ obtained as a function of the minimum lattice size used in the fit, $L_{\text{min}}$. The fit is stable from low lattice sizes, but if very small system sizes are included, there is the concern that further corrections to scaling beyond those considered in our fit could become important.  On the other hand, using a large $L_{\text{min}}$ gives larger statistical errors.  In practice, we find that the extracted value of $g_c$ is not sensitive to $L_{\text{min}}$ over a broad range, and we take the value at $L_{min}=10$ as our estimate for $g_c$.  Other RG stages show similar behavior. Table \ref{tab:CriticalCouplingValues} summarizes our final estimate of the critical coupling for each of the RG stages.  In the table, RG0 represents the tensor network for the original model, Eq.~(\ref{eq:bareT}), {\em after} truncation to 9 states (corresponding to $N_b^1 \le 2$ and $N_b^2\le 2$, see Sec.~\ref{subsec:truncationSec}), which slightly shifts $g_c$ relative to the XY model.
\begin{table}
    \caption{\label{tab:CriticalCouplingValues} Values of critical coupling $g_c$ and $\chi^2$ obtained from the fits. $\chi^2$ is shown as $\chi^2$/(number of fitting points -  number of fitting parameters).}
    \begin{ruledtabular}
    \begin{tabular}{cll}
        RG stage  & \multicolumn{1}{c}{$\chi^2$} & \multicolumn{1}{c}{$g_c$}  \\ \hline
        RG0   & $138/(160 - 17) = 0.96$ & $0.22716829(10)$ \\
        RG1   & $155/(160 - 17) = 1.1$ & $0.227116451(29)$ \\
        RG3   & $137/(160 - 17) = 0.96$ & $0.2267421500(54)$ \\
        RG5   & $163/(192 - 17) = 0.93$ & $0.22659720036(50)$
    \end{tabular}
    \end{ruledtabular}
\end{table}

We found that the critical coupling is slightly decreasing when we coarse grain. This shift is an artifact of the truncation, which generates a perturbation in the relevant parameters of the model \cite{Ueda2020FiniteScalingAnalysisBKT,Ueda2017CriticalBehaviorIcosahedron,Ueda2014DoublingEntanglementSpectrum,Ueda2023Finite-sizeBondDimensionEffectsTNR,Tagliacozzo2008ScalingEntanglementMPS,Pollmann2009TheoryFiniteEntanglementScaling,Calabrese2008Entanglementspectrum,Pirvu2012MatrixProductStateFiniteSize}. To stay at criticality, we shift the coupling $g$ for RG1, RG3 and RG5 to their respective critical couplings.

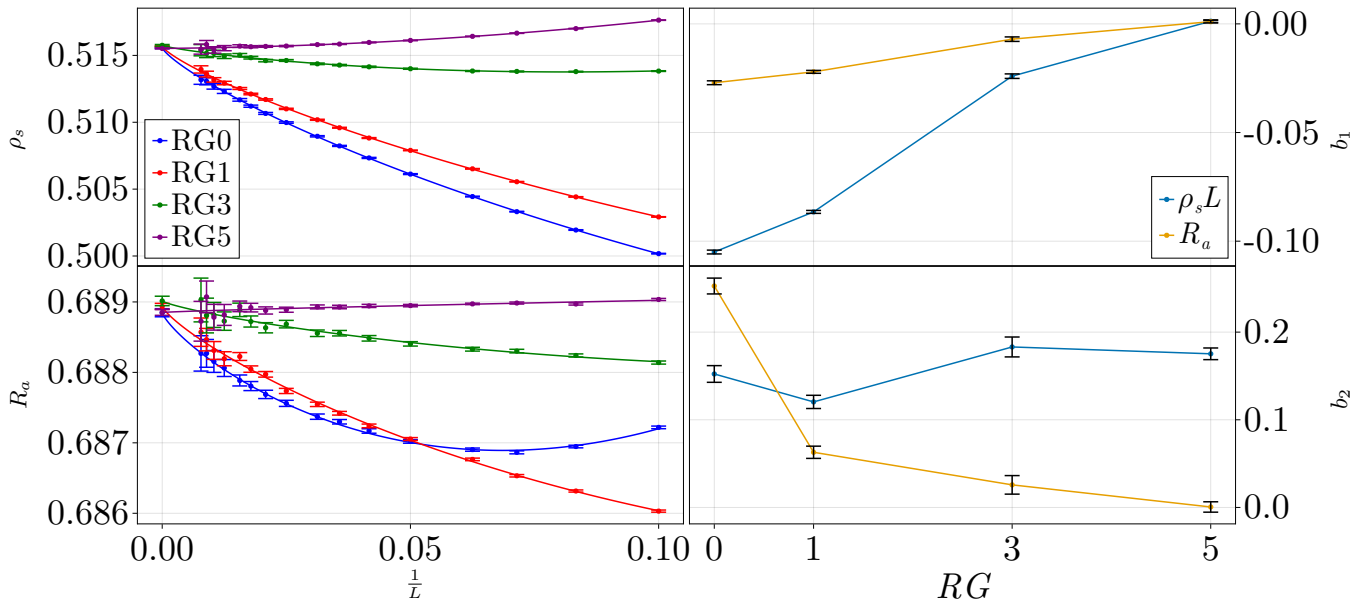
\begin{figure*}
    \begin{overpic}[width=.96\textwidth]{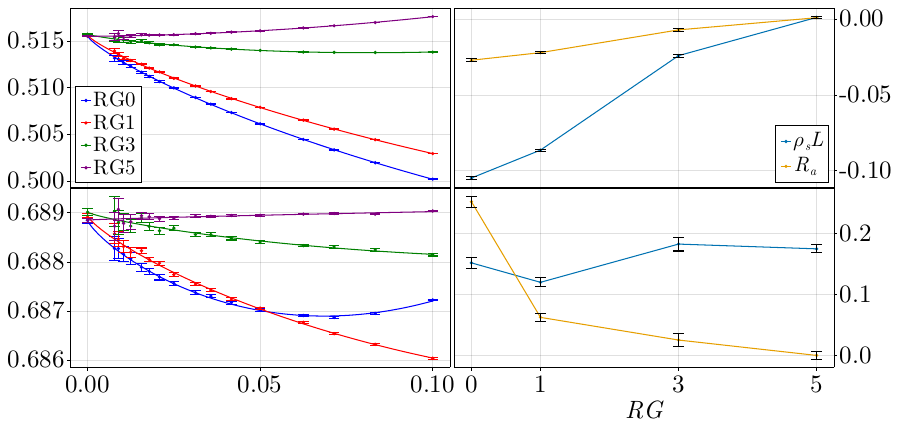}
    \put(-2, 36){\rotatebox{90}{$\rho_s$}}
    \put(-2, 16){\rotatebox{90}{$R_a$}}
    \put(100, 36){\rotatebox{90}{$b_1$}}
    \put(100, 16){\rotatebox{90}{$b_2$}}
    \put(28.5, 2){$\frac{1}{L}$}
    \end{overpic}
    \caption{\label{fig:DimensionlessObservablesCrit} On the left panels, superfluid stiffness $\rho_s L$  and wrapping probability $R_a$ as a function of $\frac{1}{L}$ for different RG stages, evaluated at the critical point. Lines are fitted with $2$ corrections to scaling, according to Eq. \eqref{eq:ScalarFitForm}. The points at $\frac{1}{L}=0$ are the universal value $R^*$ according to the fits. On the right panels, corrections to scaling as a function of number of RG steps from the values of $b_1$ and $b_2$ used in the fits. 
    The leading correction to scaling, $b_1$, is strongly suppressed for both observables.}
\end{figure*}

\subsubsection{\label{subsubsec:DimensionlessObservables} Dimensionless Observables at the Critical Point}

To view the effect of the coarse-graining on corrections to scaling, we consider observables at the critical point.
Our dimensionless observables are the superfluid stiffness times lattice size $\rho_s L$ and the three wrapping probabilities $R_x,R_2,R_a$, described in Eqs. \eqref{eq:rhos}, \eqref{eq:Rx}, \eqref{eq:Ra}, and \eqref{eq:R2}. We interpolated their values at the critical coupling from the measurements around it, and fit them to a form with two corrections to scaling:
\begin{equation}\label{eq:ScalarFitForm}
    R(g_c,L) = R^* + b_1 L^{-0.789} + b_2 L^{-2+\eta}
\end{equation}
where $R^*=\lim_{L\rightarrow\infty}{R(g_c,L)}$ is the universal value of $R$ at the critical point, which depends on the observable, and $b_1,b_2$ are the amplitudes of the two leading corrections to scaling. We again used the value $\eta=0.03810(8)$ \cite{Hasenbusch2019MonteCarloClock}.

The results for the superfluid stiffness times lattice size $\rho_s L$  and the wrapping probability $R_a$ are displayed in Fig.~\ref{fig:DimensionlessObservablesCrit}.%
The graphs are seen to become flatter with coarse-graining, indicating reduced corrections to scaling.  The correction to scaling amplitudes and values of $R^*$ are displayed on Table~\ref{tab:correctionsR} and Fig.~\ref{fig:DimensionlessObservablesCrit}.  The leading correction amplitude $b_1$ is seen to be strongly suppressed with coarse-graining.  The second correction $b_2$ is significantly suppressed for $R_a$ but not for $\rho_s L$.  This suggests that $R_a$ is less sensitive to the truncation errors than $\rho_s L$. For both observables, the values of $R^*$ agree across all RG stages. Results for the other observables are available in Appendix \ref{App:OtherObservables}.

\begin{table*}
    \begin{ruledtabular}
    \begin{tabular}{c|lll|lll|rr|rr}
                 &  \multicolumn{3}{c|}{\, $\rho_s L$} & \multicolumn{3}{c|}{$R_a$}  &  \multicolumn{2}{c|}{$G_{\rho_s L}$} & \multicolumn{2}{c}{$G_{R_a}$}  \\
           & \multicolumn{1}{c}{$R^*$} & \multicolumn{1}{c}{$b_1$} & \multicolumn{1}{c|}{$b_2$} & \multicolumn{1}{c}{$R^*$} & \multicolumn{1}{c}{$b_1$} & \multicolumn{1}{c|}{$b_2$}
            & \multicolumn{1}{c}{$b_1/R^*$} & \multicolumn{1}{c|}{$b_2/R^*$} & \multicolumn{1}{c}{$b_1/R^*$} & \multicolumn{1}{c}{$b_2/R^*$}\\ \hline
        RG0   & $0.51557(5)$ & $-0.1050(9)$ & $0.152(10)$ 
        & $0.68885(6)$ & $-0.0271(9)$ & $0.253(9)$ 
        & $-0.184(6)$ & $-0.11(6)$ 
        & $0.100(6)$ & $-1.10(6)$ \\
        RG1   & $0.51568(4)$ & $-0.0865(7)$ & $0.120(7)$ 
        & $0.68894(4)$ & $-0.0221(7)$ & $0.063(7)$
        & $-0.149(4)$ & $0.47(4)$ 
        & $0.080(4)$ & $-0.07(4)$
        \\
        RG3   & $0.51575(6)$ & $-0.0241(10)$ & $0.183(11)$ 
        & $0.68901(7)$ & $-0.0070(10)$ & $0.026(10)$ 
         & $-0.027(6)$ & $0.77(6)$ 
         & $0.036(6)$ & $0.14(6)$
         \\
        RG5   & $0.51552(4)$ & $\phantom{-}0.0012(6)$ & $0.175(7)$ 
        & $0.68885(4)$ & $\phantom{-}0.0010(6)$ & $0.001(6)$ 
        & $0.018(3)$ & $0.81(3)$ 
        & $0.009(4)$ & $0.24(4)$
    \end{tabular}
    \end{ruledtabular}
    \caption{\label{tab:correctionsR} Corrections to scaling.  Values of $R^*$, $b_1$ and $b_2$ used in the fits of the dimensionless observables to Eq.~\eqref{eq:ScalarFitForm} in Fig.~\ref{fig:DimensionlessObservablesCrit}, and values of $b_1/R^*$ and $b_2/R^*$ used in the fits of the derivative observables to Eq.~\eqref{eq:DerivativeFittingForm} in Fig.~\ref{fig:DerivativeObservablesCrit}.
    }
\end{table*}

For both observables displayed, the fits of the different RG stages agree about the universal value of $R^*$ at $L\rightarrow\infty$, indicating that the RG procedure indeed leaves us within the same universality class. In addition, we see that the graphs are flattening as we coarse grain, indicating a reduction in corrections to scaling. 

Note that were it not for the truncation, the graph of RG1 would exactly match that of RG0, but stretched horizontally by a factor of $2$.  The corrections to scaling amplitudes would then reduce by a factor $2^{-\omega_m}$ for each coarse-graining step. We don't observe these behaviors since the truncation effectively adds perturbations (both relevant and irrelevant) to the model.  These perturbations are expected to be reduced with increasing bond dimension $D$.

\subsubsection{\label{subsubsec:ObservableDerivatives} Derivatives of Dimensionless Observables}

For the derivatives, we use a different fitting form that takes into account their scaling dimension:
\begin{equation}\label{eq:DerivativeFittingForm}
     G_R(g_c,L)L^{-\frac{1}{\nu}} = 
     R^* + b_1 L^{-0.789} + b_2 L^{-2+\eta}
\end{equation}
where we take $\nu=0.67169(7)$ and $\eta=0.03810(8)$ \cite{Hasenbusch2019MonteCarloClock} and fit for $3$ parameters: $R^*$, the critical point value at the thermodynamic limit, and $b_1,b_2$, the corrections to scaling coefficients.  Note that, since $G_R$ is a derivative with respect to $g$, the value of $R^*$ is not universal, since it depends on the units of $g$.  In particular, the value of $R^*$ depends on the RG stage used.  Therefore, when comparing different RG stages, it is most meaningful to consider $\frac{G_R\cdot L^{-\frac{1}{\nu}}}{R^*}$.

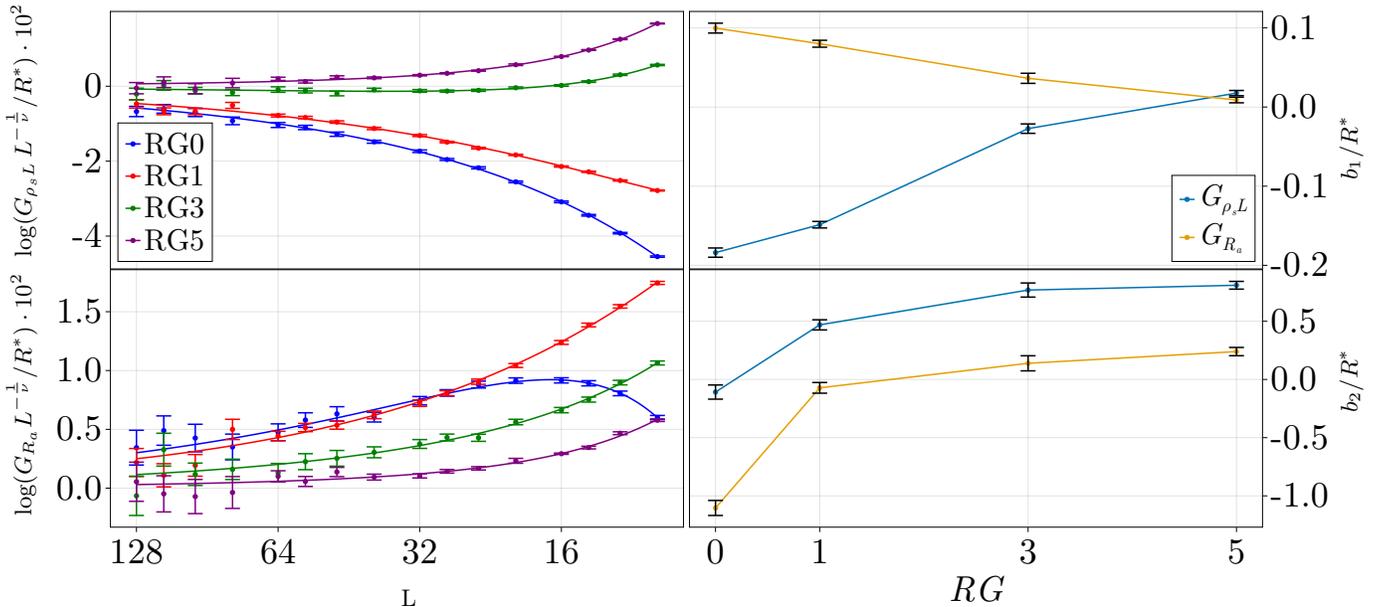
\begin{figure*}
    \begin{overpic}[width=.96\textwidth]{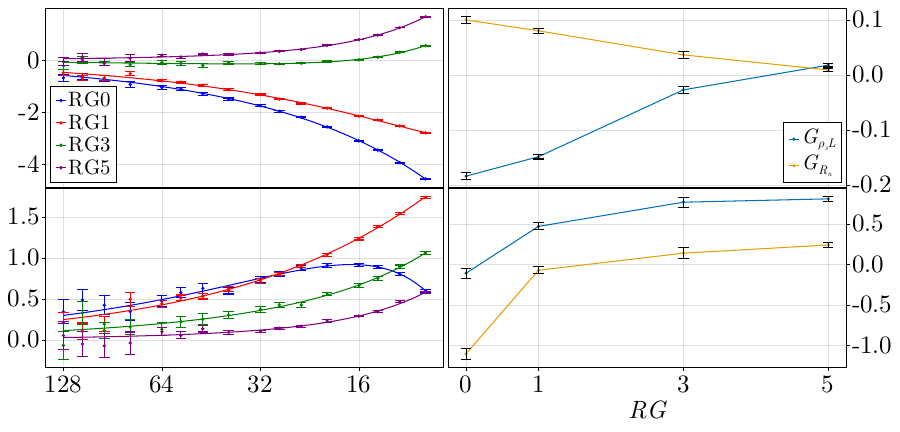}
    \put(-3, 28){\rotatebox{90}{$\log (G_{\rho_s L}\, L^{-\frac{1}{\nu}}/R^*) \cdot 10^2$}}
    \put(-3, 8){\rotatebox{90}{$\log (G_{R_a} L^{-\frac{1}{\nu}}/R^*) \cdot 10^2$}}
    \put(27.5, 1){L}
    \put(100, 34){\rotatebox{90}{$b_1/R^*$}}
    \put(100, 16){\rotatebox{90}{$b_2/R^*$}}
    \end{overpic}
    \caption{\label{fig:DerivativeObservablesCrit} On the left panels, normalized derivatives of $\rho_s L$ and $R_a$ as a function of $L$ for different RG stages, evaluated at the critical point.  The derivatives have been rescaled by $L^{-1/\nu}$ with $\nu=0.67169(7)$ \cite{Hasenbusch2019MonteCarloClock}.  The plot is in log-log scale, such that a slope represents a correction to the cited critical exponent. Lines are fitted according to Eq.~\eqref{eq:DerivativeFittingForm}.  On the right panels, corrections to scaling from the values of $b_1/R^*$ and $b_2/R^*$ used in the fits.  The leading correction to scaling, $b_1/R^*$, is strongly suppressed for both observables.}
\end{figure*}

The derivatives of the superfluid stiffness times lattice size, $G_{\rho_s L}$, and wrapping probability, $G_{R_a}$, are displayed in Fig.~\ref{fig:DerivativeObservablesCrit}.%
The figures are on a log-log scale, so the slope is equivalent to a correction to the critical exponent $\nu$ relative to the cited value. The amplitudes of the corrections to scaling are displayed in Table~\ref{tab:correctionsR} and Fig.~\ref{fig:DerivativeObservablesCrit}. As before, we observe that the leading correction to scaling $b_1/R^*$ is systematically suppressed by the coarse-graining, whereas the second correction $b_2/R^*$ is suppressed for $G_{R_a}$ but not for $G_{\rho_s L}$.  Results for the other observables are available in Appendix~\ref{App:OtherObservables}.

\subsubsection{\label{subsubsec:CriticalExponentNu} Extraction of Critical Exponent $\nu$}

We now extract our best estimate of the critical exponent $\nu$. We use the results of RG5 with a fitting form excluding the first correction to scaling (assuming the first correction to scaling is sufficiently suppressed by the RG; including it yields similar results with larger error bars):
\begin{equation}\label{eq:NuFittingForm}
    G_R(g_c,L) = L^{\frac{1}{\nu}} (R^* + b_2 L^{-2.0+\eta} + b_3 L^{-2.02})
\end{equation}
where we now have four fitting parameters: $\nu$, $R^*$, $b_2$ and $b_3$.  We perform a fit on all lattice sizes, up to a minimal size used in the fit $L_{\mathrm{min}}$.

\begin{figure}
    \includegraphics{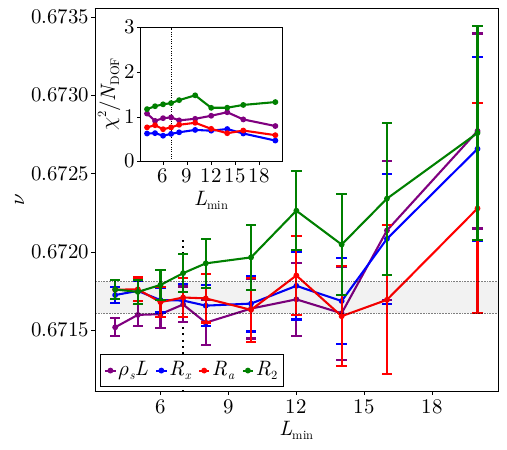}
    \caption{\label{fig:NuExtraction} Critical exponent $\nu$ extracted from $G_{\rho_s L},$ $G_{R_x},$ $G_{R_a},$ $G_{R_2}$ vs minimal lattice size used for the fit $L_{min}$ using the RG5 model. Inset: $\chi^2$ divided by number of degrees of freedom vs minimal lattice size used for the fit. The dotted vertical line at $L_{min}=7$ in the main figure and in the inset indicates the point we used for our final estimate of $\nu$, displayed by the shaded region.}
\end{figure}

The extracted critical exponent $\nu$ vs $L_\mathrm{min}$ is displayed in Fig.~\ref{fig:NuExtraction}.%
In the inset, $\chi^2$ divided by the number of degrees of freedom is shown as a measure of goodness of fit. We see that starting from $L_\mathrm{min}=6$ the extracted $\nu$ is consistent for all the observables.  This may be an indication that below $L_\mathrm{min}=6$ further corrections to scaling, beyond Eq.~\eqref{eq:NuFittingForm}, would be needed in the fit to make it unbiased.

For the final estimate of the critical exponent we perform a fit using all of the observables ($\rho_S L$, $R_x$, $R_a$, and $R_2$) at once for $L_{min}=7$ and obtain $\nu=0.67171(10)$. Our estimate is consistent with $\nu=0.67169(7)$ of Ref.~\onlinecite{Hasenbusch2019MonteCarloClock}.

\subsubsection{\label{subsubsec:CriticalExponentEta} Extraction of Critical Exponent $\eta$}

The closing time $T_\mathrm{w}$, defined in Eq.~\eqref{eq:Tw}, is a susceptibility-like observable that allows us to extract the critical exponent $\eta$ by fitting it to the form:
\begin{equation}\label{eq:EtaFittingForm}
    T_\mathrm{w}(g_c,L) = L^{2-\eta} (T_\mathrm{w}^* + b_2 L^{-2.0+\eta} + b_3 L^{-2.02})
\end{equation}
where we again skip the first correction to scaling, assuming that for RG5 the leading correction to scaling is negligible. We again have four fitting parameters: $\eta$, $T_\mathrm{w}^*$, $b_2$ and $b_3$.

\begin{figure}
    \includegraphics{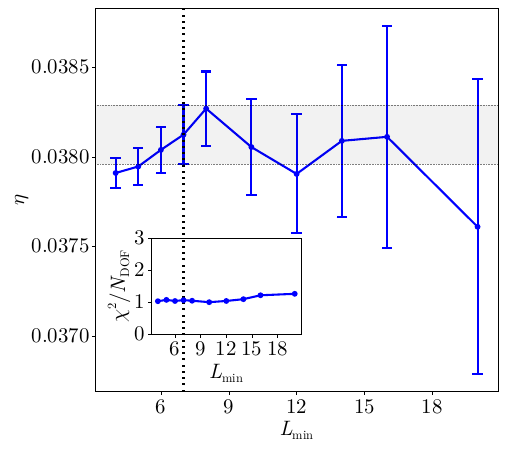}
    \caption{\label{fig:EtaExtraction} Critical exponent $\eta$ extracted from $T_\mathrm{w}$ vs minimal lattice size used for the fit $L_{min}$ using the RG5 model. Inset: $\chi^2$ divided by number of degrees of freedom vs minimal lattice size used for the fit. The dotted vertical line at $L_{min}=10$ in the main figure and in the inset indicates the point we used for our final estimate of $\eta$, diplayed by the shaded region.}
\end{figure}

The extracted critical exponent $\eta$ vs minimal lattice size used for the fit $L_\mathrm{min}$ is displayed in Fig.~\ref{fig:EtaExtraction}.%
The inset shows $\chi^2$ divided by the number of degrees of freedom. We see that even starting from $L_\mathrm{min}=4$ the fit is consistent with the data.

As a conservative final estimate of $\eta$ we take the value for $L_\mathrm{min}=7$, same as in our estimate of $\nu$, and obtain $\eta=0.03813(16)$. Our estimate is consistent with $\eta=0.03810(8)$ of Ref.~\onlinecite{Hasenbusch2019MonteCarloClock}.

\subsection{\label{subsec:2ptFunc} Two-Point Function}

The two-point function is defined as:
\begin{equation}
    c(\mathbf{r}) = \langle \psi_i \psi_{i+\mathbf{r}} \rangle
\end{equation}
where $\mathbf{r}$ is a distance on the lattice. We measure it by counting the number of times an open worm spans a distance $\mathbf{r}$ in the Monte Carlo \cite{Prokofev2001WormAlgorithms}. In the coarse-grained models we use the edge tensors described in Sec.~\ref{subsec:TensorMonteCarlo} which decimate the edge of the tensor to a specific site within the coarse-grained supersite. Thus, the two-point function measured on a coarse-grained model is the decimated one: $c_{\text{RG1}}(\mathbf{r})=c_{\text{RG0}}(2\mathbf{r})$. For more details, see App.~\ref{app:MCmethods}.

We look for two improvements to the two-point function when we coarse grain. First, the coarse-grained model allows us to simulate the two-point function of a large lattice using a smaller lattice, thus saving computational resources.  Second,  since coarse-grained supersites contain several internal sites, we expect the coarse-grained model to be less sensitive to the lattice, {\em i.e.} to be closer to the continuum model.  For instance, as a manifestation of this, we expect the rotational symmetry broken by the lattice to be partially restored.

\begin{figure}
    \includegraphics[width=0.9\linewidth]{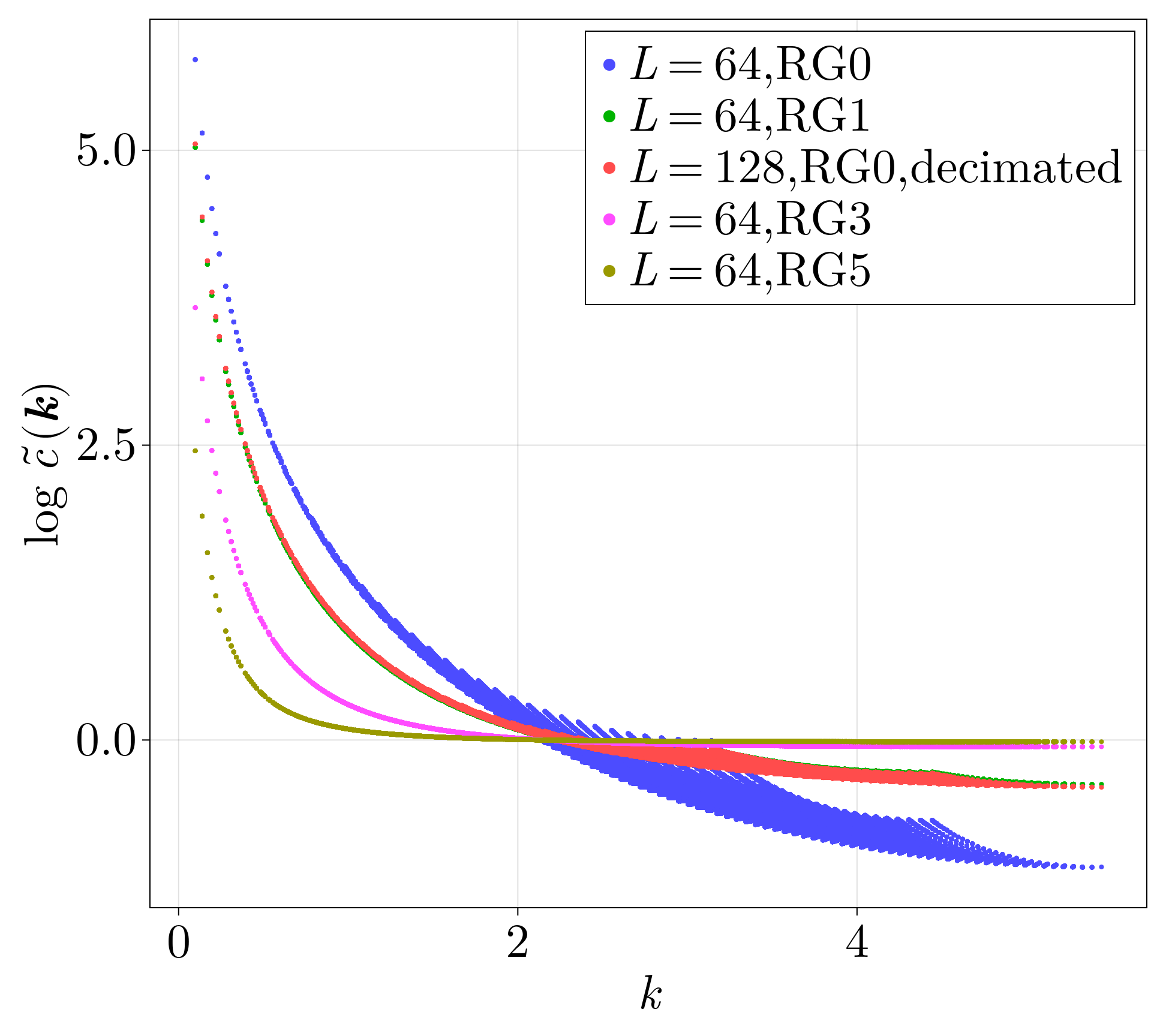}
    \caption{\label{fig:2ptMomentum} Log momentum space two-point function $\log\Tilde{c}(\mathbf{k})$ as a function of log absolute value of the momentum $\log |\mathbf{k}|$, on a cubic lattice of size $L=64$ using all RG stages, and using RG0 with lattice size $L=128$ decimated to $L=64$. Points from all the 3D momentum space are displayed.}
\end{figure}

Figure~\ref{fig:2ptMomentum}%
compares the results of a system on lattice size $L=64$ using the RG1 model with a system on lattice size $L=128$ using the RG0 model. The two-point function for $L=128$ is decimated to size $L=64$ by keeping only even positions on the real-space lattice. The points match relatively well without the need for rescaling or other fitting parameters, indicating that we succeed in obtaining results of a $L=128$ lattice while simulating a lattice of smaller size $L=64$ on a coarse-grained model.

In Fig.~\ref{fig:2ptMomentum} we also show the two-point function in momentum space for all RG stages on a cubic lattice of size $L=64$. The splay of the points at large momenta is a result of corrections to scaling due to the lattice breaking rotational symmetry. The splay in the points is smaller for RG1 compared to RG0, indicating that rotational symmetry is partially restored. The splay keeps decreasing as we coarse grain further.

\begin{figure}
    \includegraphics{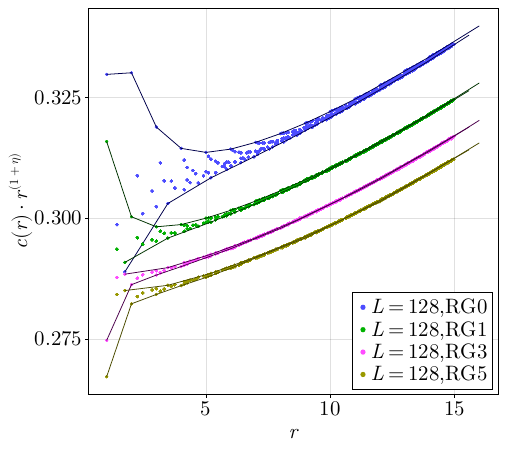}
    \caption{\label{fig:2ptRealspace} Real space two-point function $c(r)$ multiplied by $r^{1+\eta}$ as a function of distance $r$, on a cubic lattice of size $L=128$ for all RG stages. All points from the 3D space are displayed. The graphs were independently rescaled to be shown on the same graph, with the normalizations chosen arbitrarily. Error bars are smaller than the size of the points. The lines follow the points along the principal axes $(i,0,0)$ and points on the diagonal $(i,i,i)$ for each RG stage.}
\end{figure}

It is also instructive to look at the two-point function in real space, Fig.~\ref{fig:2ptRealspace}.%
The expected behavior is 
\begin{eqnarray}
c({\mathbf{r}})={\mathcal{A}}\, r^{-1-\eta}\, \Phi({\mathbf{r}}/L),
\end{eqnarray} 
up to non-universal corrections that are expected at short distances. Here,  ${\mathcal{A}}$ is a non-universal amplitude and $\Phi({\mathbf{r}}/L)$ is a universal function that reflects the sensitivity of the correlation function to the finite extent of the system.  As $L$ grows, for any given fixed $r$, this function becomes increasingly flat as it approaches $\Phi(0)$. In Fig.~\ref{fig:2ptRealspace} we multiply the two-point function $c(r)$ by $r^{1+\eta}$ to remove the power law and compare the residual. We used $\eta=0.03810(8)$ \cite{Hasenbusch2019MonteCarloClock}. Since the overall scale of the two-point function is non-universal (and therefore dependent on the RG stage) we rescale the two-point function for each RG stage by an arbitrary factor to display the results together.  

The residual in Fig.~\ref{fig:2ptRealspace} has a weak $r$ dependence at large distances, from which it is possible to extract the universal function  $\Phi({\mathbf{r}}/L)$. At short distances, on the other hand, non-universal corrections arise from the discreteness of the lattice.  These lead to a splay in the points at short distances.  The lines in Fig.~\ref{fig:2ptRealspace} follow the two-point function on the principal axis $(i,0,0)$ and the diagonal $(i,i,i)$.  The splay is largest for RG0 and, as we coarse-grain, the splay reduces. For instance, for RG3, the splay at $r=4$ is $0.25 \%$ of the average value.  To obtain a comparably small splay at RG0, it is necessary to go to $r=15$.  This shows that the coarse-grained model has an emergent spherical symmetry at a much smaller scale, indicating a much weaker sensitivity to the discreteness of the lattice.

\section{\label{sec:methods} Methods}

In this section, we expand on the discussion in Sec.~\ref{sec:RGImproved} and present our method in detail.  For notational simplicity, we first present a two-dimensional version of the method on the square lattice.  We will later explain how to generalize the method to three-dimensional models on the cubic lattice.

\subsection{\label{subsec:generalMethod} Coarse-graining Method}

In this section, we explain how we coarse-grain to obtain an improved action for Monte Carlo simulations.  The starting point of our method is a partition function that is represented as a tensor network on the square lattice, see Eq.~\eqref{eq:tensorZ} and Fig.~\ref{fig:T}.
We assume that all the tensor elements $T_{x x^\prime y y^\prime}$ are non-negative.  This ensures that the partition function can be importance-sampled using Monte Carlo methods without a sign problem.

\begin{figure}
    \includegraphics{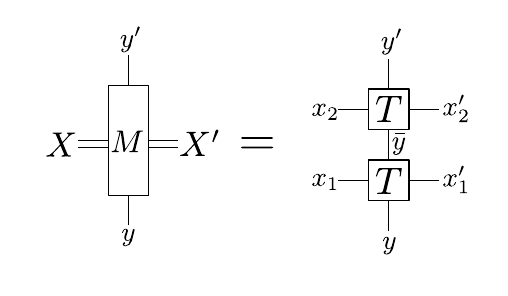}
    \caption{\label{fig:TTtoM} Coarse graining two tensors by contracting over the common bond $\bar{y}$, as in Eq.~\eqref{eq:coarsegrain}. Double lines and single lines denote bonds with bond dimensions $D^2$ and $D$ respectively.}
\end{figure}

We coarse-grain in one direction at a time, alternating between the $y$ and $x$ directions.   We only describe the step in the $y$ direction, since the step in the $x$ direction is the same up to rotation of the lattice.  The coarse-graining step consists of contracting two neighboring tensors over their common bond index:
\begin{equation}
    M_{X X^\prime y y^\prime} = \sum_{\bar{y}=1}^D {T_{x_1 x_1^\prime y \bar{y}} T_{x_2 x_2^\prime \bar{y} y^\prime}}
    \label{eq:coarsegrain}
\end{equation}
where the indices are as shown in Fig.~\ref{fig:TTtoM},
and where we introduced the composite bond indices $X=(x_1,x_2)$ and $X^\prime=(x_1^\prime,x_2^\prime)$.  Since these indices run over $D^2$ values, the coarse-grained tensor $M$ has a higher dimension in the $x$ direction compared to $T$.  We next describe how we truncate the tensor to constrain the bond dimension.

To truncate, we begin by defining a tensor $S$ 
\begin{equation}
    S_{X X^\prime y y^\prime} = \sum_{\bar{X},\bar{X}^\prime=1}^{D^2}{
        M_{\bar{X}\bar{X}^\prime yy^\prime}
        O_{\bar{X},X}
        O_{\bar{X}^\prime,X^\prime}
    }\label{eq:Sdef}
\end{equation}
where $O_{\bar{X},X}$ is a $D^2\times D^2$ orthogonal matrix to be chosen later.  This amounts to a change of basis for the compound indices $X$ and $X^\prime$. If we restrict $X$ and $X^\prime$ to the first $D$ values, we obtain
\begin{equation}
    \tilde{T}_{x x^\prime y y^\prime} = \sum_{\bar{X},\bar{X}^\prime=1}^{D^2}{
        M_{\bar{X}\bar{X}^\prime yy^\prime}
        \tilde{O}_{\bar{X},x}
        \tilde{O}_{\bar{X}^\prime,x^\prime}
    }\label{eq:Ttilde}
\end{equation}
where the restricted tensor $\tilde{T}_{x x^\prime y y^\prime} = S_{x x^\prime y y^\prime}$ has the same dimension as the original $T$ tensor, and $\tilde{O}_{\bar{X}, x} = O_{\bar{X}, x}$ is a $D^2\times D$ orthogonal matrix.  Throughout we follow the convention in which upper case indices run over $D^2$ values and lower case indices over $D$ values.

\begin{figure}
    \includegraphics{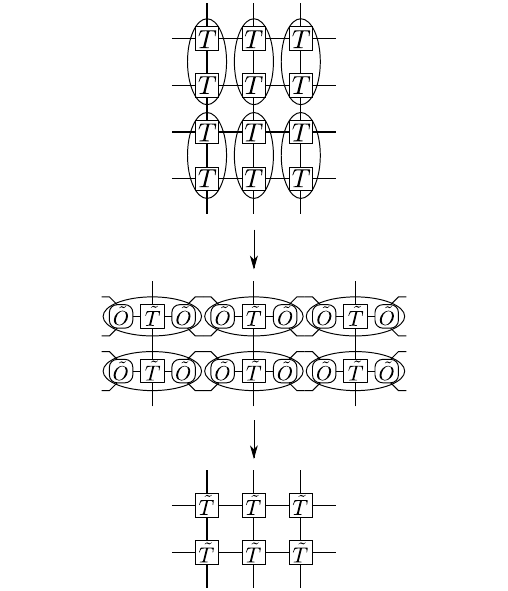}
    \caption{\label{fig:HOTRG} Depiction of the coarse-graning method. Starting from the tensor network, we contract inside supersites, truncate the coarse-grained tensors, and obtain the new tensor network.}
\end{figure}

We now define a tensor $\tilde{M}$,
\begin{equation}
    \Tilde{M}_{X X^\prime y y^\prime} = 
    \sum_{x,x^\prime = 1}^{D}{
        \Tilde{T}_{xx^\prime y y^\prime}
        \Tilde{O}_{X,x}\Tilde{O}_{X^\prime,x^\prime}
    }
\end{equation}
by transforming the truncated tensor $\tilde{T}$ back to the original basis.  This acts as an approximation for $M$.  Our goal will be to choose $\tilde{O}$ to maximize the truncation fidelity, defined as:
\begin{align}\label{eq:TruncationFidelity}
 \mathcal{F} =& 1-\frac
    {\sum_{X X^\prime=1}^{D^2}{\sum_{y y^\prime =1}^D{
        \left(M_{X X^\prime y y^\prime}-\tilde{M}_{X X^\prime y y^\prime}\right)^2
    }}}
    {\sum_{X X^\prime=1}^{D^2}{\sum_{y y^\prime =1}^D{
        \left(M_{X X^\prime y y^\prime}\right)^2
    }}} \\
    =& \frac
    {\sum_{x x^\prime=1}^{D}{\sum_{y y^\prime =1}^D{
        \left(\Tilde{T}_{xx^\prime yy^\prime}\right)^2
    }}}
    {\sum_{X X^\prime=1}^{D^2}{\sum_{y y^\prime =1}^D{
        \left(M_{X X^\prime y y^\prime}\right)^2
    }}}\, .
\end{align}
When we replace $M$ by $\tilde{M}$ in the tensor network, the orthogonal matrices $\Tilde{O}$ of neighboring sites contract into the identity $\sum_{X=1}^{D^2}{\Tilde{O}_{X,x}\Tilde{O}_{X,x^\prime}}=\delta_{x,x^\prime}$ and we obtain a coarse-grained lattice with the restricted tensors $\Tilde{T}$, see Fig.~\ref{fig:HOTRG}.%

One method to choose $\tilde{O}$ is by using the higher-order singular value decomposition (HOSVD)\cite{DeLathauwer2000HOSVD,Xie2012RGbyHOSVD}. HOSVD is a decomposition similar to the singular value decomposition (SVD), but on higher-order tensors instead of matrices. It consists of rearranging the indices of the $M$ tensor to a matrix:
\begin{equation}\label{eq:Mprime}
    M^\prime_{X,X^\prime y y^\prime} =
    M_{X X^\prime y y^\prime}
\end{equation}
where we view $M^\prime$ as a 2D matrix with indices $X$ and $X^\prime y y^\prime$. Then the positive symmetric matrix $M^\prime M^{\prime T}$ is diagonalized:
\begin{equation}\label{eq:spectralDecomposition}
    M^\prime M^{\prime T}= O \Lambda O^T\,,
\end{equation}
where $\Lambda$ has non-negative diagonal entries ordered from largest to smallest, whose values are
\begin{equation}
\Lambda_X = \sum_{X^\prime=1}^{D^2}\sum_{y,y^\prime=1}^D{|S_{XX^\prime yy^\prime}|^2}\,.    
\end{equation}
Here, $S$ is the tensor defined in Eq.~\eqref{eq:Sdef} with the orthogonal matrix $O$ obtained in Eq.~\eqref{eq:spectralDecomposition}.  Hence, we see that by keeping the $D$ largest eigenvalues, we are maximizing the spectral weight in the truncated tensor.  Finally, we obtain $\tilde{O}$ by restricting $O$ as described above.

The HOSVD based truncation scheme doesn't necessarily maximize the truncation fidelity, but it gives a good approximation nonetheless \cite{DeLathauwer2000HOSVD}.  However, the HOSVD is expected to generate renormalized tensors $\tilde{T}$ with negative values, as we explain in Sec.~\ref{subsec:SignProblemHOSVD} below. Tensors with negative values are not suitable for Monte Carlo simulations due to the sign problem, which severely degrades the convergence rate of observables in the simulation.  Therefore, we cannot use the HOSVD directly. 

Instead, we introduce a different method, the {\em Positive Tensor Truncation} (PTT), that yields a truncated tensor  $\Tilde{T}$  that is guaranteed to be non-negative.   Later, we will apply this method to the $XY$ model and demonstrate that its fidelity is similar to HOSVD.

\begin{figure}
    \includegraphics{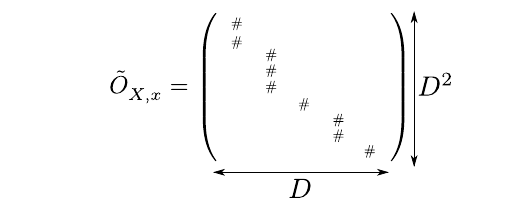}
    \caption{\label{fig:U} Structure of the orthogonal non-negative matrix $\Tilde{O}_{X,x}$. In the matrix, $\#$ stands for positive values while empty spaces represent zero entries.}
\end{figure}

The main idea behind PTT is to choose $\Tilde{O}$ to be non-negative. The original tensor $T$ is itself non-negative and therefore, by Eq.~\eqref{eq:Ttilde}, this choice suffices to ensure that $\tilde{T}$ has the same property.  This places severe constraints on the form of $\tilde{O}$, whose columns are orthonormal vectors.  For these vectors to be non-negative, they must be non-overlapping, see Fig.~\ref{fig:U}%

The PTT method consists of partitioning the $D^2$ bond states into $D$ separate subsectors.  The choice of how to partition will be discussed when we present a concrete example, the XY model.  Each subsector yields one column of $\Tilde{O}$. To obtain this column, we restrict the matrix $M^\prime M^{\prime T}$ to the subsector and diagonalize it.  We then select the state with the largest eigenvalue. This maximizes the truncation fidelity within the subsector and guarantees that the column is non-negative -- since $M^\prime M^{\prime T}$ is positive and symmetric, the vector with the largest eigenvalue can always be chosen to be non-negative.  Thus, we keep a single state for each subsector, which we dub the {\em superstate} of that subsector.

In three dimensions, just as in two dimensions, we coarse grain in one direction at a time, alternating between the $y$, $z$, and $x$ directions. However, unlike 2D, each time that we coarse grain, two transverse directions must be truncated \cite{Xie2012RGbyHOSVD}.  For instance, when we coarse-grain in the $y$ direction, in analogy to Eq.~\eqref{eq:coarsegrain}, we obtain a tensor $M_{XX^\prime y y^\prime Z Z^\prime}$, where $Z$ and $Z^\prime$ take $D^2$ values.  To truncate, we introduce two distinct orthogonal tensors $\tilde{O}^{(X)}_{X,x}$ and $\tilde{O}^{(Z)}_{Z,z}$.  To obtain  $\tilde{O}^{(X)}_{X,x}$ we proceed as above, replacing Eq.~\eqref{eq:Mprime} by $M^\prime_{X,X^\prime y y^\prime Z Z^\prime}=M_{XX^\prime y y^\prime Z Z^\prime}$ (and similarly for $\tilde{O}^{(Z)}_{Z,z}$, but with $M^\prime_{Z,XX^\prime y y^\prime Z^\prime}=M_{XX^\prime y y^\prime Z Z^\prime}$).  
After we finish coarse-graining in all three directions the resulting tensor is anisotropic, due to the order of the coarse-graining sequence.  We symmetrize the tensor at this point, to ensure that it has the cubic point group symmetry.  This completes a single full coarse-graining step.

\subsection{Truncation of states}

\label{subsec:truncationSec}

Current conservation at each site implies that the net current (ingoing minus outgoing) on a bond is a good quantum number to label states in the tensor representation. Current conservation is both a necessary and sufficient condition for O$(2)$ symmetry.  Hence, we must maintain exact current conservation at each stage of the truncation to ensure that the truncated model remains in the $XY$ universality class.

In the bare model, without coarse-graining, the states of the bonds are determined by the two bond numbers $N_b^1,N_b^2$. To obtain a tensor with a finite number of elements that we can store and use for simulations, we truncate to states with $N_b^1 \leq 2$ and $N_b^2 \leq 2$, leaving a total of $3 \times 3 = 9$ possible states for each bond. We choose to work in the hard spin limit, $\lambda\to \infty$.  Then, these states constitute more than $99.99\%$ of the states encountered during simulations at the critical point.  The tensor network after this initial truncation will be labeled RG0.

When we coarse-grain, we truncate using the PTT described in Sec.~\ref{subsec:generalMethod}.  This requires a partition of states into subsectors.  Since current is a good quantum number, all states in a subsector should have the same current.  However, beyond this constraint, it is not a priori clear how to choose subsectors.

One simple approach is to group all states with a given current into a subsector, and to keep as many subsectors as allowed by the computational resources.  For example, all the states with zero net current in $x=1$, all the states with $1$ net current in $x=2$, all the states with $-1$ net current in $x=3$, etc. We tried this approach with $D=11$ subsectors, with currents ranging from $-5$ to $+5$.  However, we found that this approach gives a very low truncation fidelity, as compared to HOSVD.  This is because states with small currents dominate the partition function at the critical point, as discussed above.  Therefore, the inclusion of states with large currents is not an efficient method of truncation.

A much better approach is to throw out large current states and allot many subsectors to the low current states.  One option is to restrict the net current to be at most $+2$, and to divide the states with net current $\pm 1$ into subsectors.  Hence, we choose a single subsector each for net current $0$, $2$, and $-2$.  Each subsector yields a single superstate.  

\begin{figure}
    \includegraphics{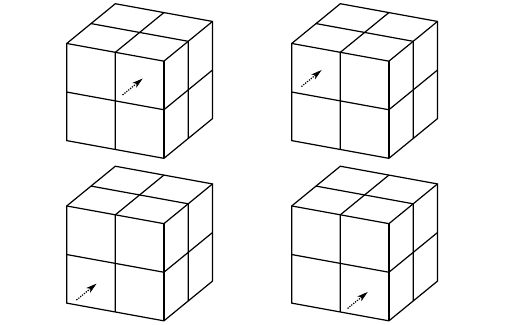}
    \caption{\label{fig:cornerstates} The corner states, where the current is concentrated in a specific corner of the face.}
\end{figure}

For the net current $+1$ we choose four subsectors.  It is natural to choose these according to where the current is concentrated among the four corners of the face, as shown in Fig.~\ref{fig:cornerstates}.%
We call these superstates ``corner states''. They contain states that have a single current going in at the designated corner and zero current at the other corners, and also other states, such as two currents coming in and one coming out, where the corner is assigned according to the position of the ``center of current'' (the average location of the current weighted by its value).    Similarly, there are four corner states with current $-1$.  In total, we used $11$ states: a single state with zero net current, $8$ corner states for net current=$\pm1$, and two states with net current=$\pm2$. See App.~\ref{app:NNHOTRG} for more details.

\begin{figure}[h]
    \includegraphics[width=0.9\linewidth]{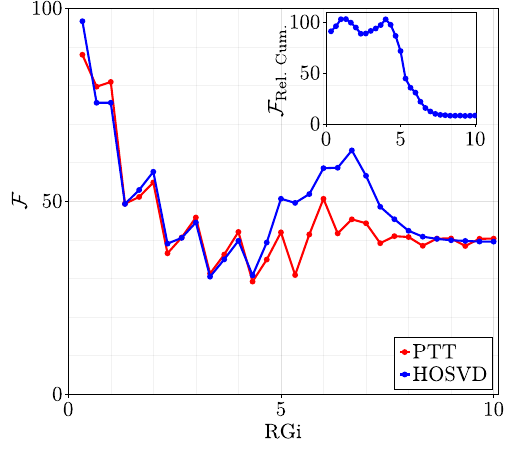}
    \caption{\label{fig:TruncationErrorPerStep} Main Plot: Truncation fidelity of the PTT and HOSVD, according to Eq.~\eqref{eq:TruncationFidelity}, both starting from RG0 at the critical point $g=0.227168$. Each coarse-graining step is made up of three substeps, in the $y$, $x$, and $z$ directions. The truncation fidelity is plotted for each substep. Inset: Relative cumulative truncation fidelity between the PTT and the HOSVD, defined in Eq.~\eqref{eq:RelativeCumulativeFidelity}.}
\end{figure}

In Fig.~\ref{fig:TruncationErrorPerStep}%
we compare the truncation fidelities of the HOSVD and the PTT using the discussed subsectors, both using a total of $D=11$ states. Both were carried out $10$ times consecutively, starting from the critical point of RG0, $g=0.227168$.

The truncation fidelity is similar between the two truncation methods, indicating that the PTT is comparable to HOSVD in terms of the quality of approximation of the coarse-grained tensor.

In the inset we show the relative cumulative truncation fidelity, 
\begin{equation}\label{eq:RelativeCumulativeFidelity}
    \mathcal{F}_{\text{Rel.~Cum.~}} = \frac{\prod_i \mathcal{F}^i_{\text{PTT}}}{\prod_i \mathcal{F}^i_\text{HOSVD}}   
\end{equation}
  which is a measure of the relative quality of the truncation methods across multiple RG steps. The plot indicates that the PTT has a similar cumulative fidelity compared to the HOSVD up to five RG steps, where it starts dropping down, approaching $8\%$ of the cumulative fidelity of the HOSVD after ten steps. This again indicates that, at least for the first few steps, the PTT performs comparably to HOSVD.

\subsection{\label{subsec:SignProblemHOSVD} Sign Problem of the HOSVD}

\begin{figure}[h!]
    \includegraphics{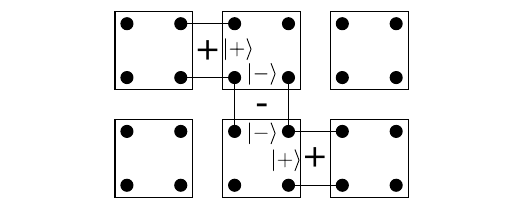}
    \caption{\label{fig:signproblem} Symmetric - anti-symmetric - symmetric kink configuration that generates a sign problem in the HOSVD.}
\end{figure}

In this section we explain why the HOSVD is expected to generate a sign problem, by considering the example of the XY model on a cubic lattice.  In this model, the states are denoted by the current going through the bonds, $|N_b^1,N_b^2\rangle$. For simplicity, we restrict ourselves to states with $N^1_b,N^2_b \leq 1$. There are two states with net current=$1$: $|1,0\rangle$ and $|0,1\rangle$. Since net current is a good quantum number, the HOSVD generates two states with net current=$1$ that respect the symmetry between the two bonds - a symmetric state and an anti-symmetric state:
\begin{eqnarray}
    |+\rangle &= \frac{1}{\sqrt{2}} (|1,0\rangle + |0,1\rangle) \\
    |-\rangle &= \frac{1}{\sqrt{2}} (|1,0\rangle - |0,1\rangle)
\end{eqnarray}
Next, we consider a kink configuration with two symmetric states around an anti-symmetric state, as shown in Fig.~\ref{fig:signproblem}.%
 Due to the symmetry of the model, the two tensors at the kink always have the same value with an opposite sign, so the configuration always generates a negative sign.
 
 Therefore, the renormalized tensor $\Tilde{T}$ obtained using the HOSVD contains negative values, rendering it unsuitable for Monte Carlo simulations due to the sign problem.

\subsection{\label{subsec:TensorMonteCarlo} Tensor Worm Monte Carlo}

In this section, we describe our Monte Carlo algorithm. We use a worm algorithm \cite{Prokofev2001WormAlgorithms}, since it allows us to reduce the effect of critical slowing down and to maintain current conservation, thus ensuring that the model is within the XY universality class. However, the algorithm is adjusted to conform with tensor weights, as we explain below.

In the worm algorithm, the physical state space, which consists of closed paths as in Eq.~\eqref{eq:Zworm}, is expanded to include states with a single open path, called the worm. Then, the worm is built by shifting one of its edges until it closes up, returning to the physical space of closed paths.

The open worm states have weights $Z^\prime(i_1,i_2)$, where $i_1,i_2$ are the start and the end of the open worm. In the standard worm algorithm, the weights are given by Eq.~\eqref{eq:Zworm}, except that the sum is over closed paths in the presence of an open worm. In the tensor network representation, we need to use special ``edge tensors'', $T^E$ on sites $i_1,i_2$ to accommodate for the lack of current conservation on these sites.  Accordingly, the probability of shifting the worm's edge is determined by the (bulk) tensor $T$ and the edge tensor $T^E$. 

When we coarse grain, we also coarse grain the edge tensors,
\begin{eqnarray}
    M^E_{X X^\prime y y^\prime} =&& \sum_{\bar{y}=1}^D {T^E_{x_1 x_1^\prime y \bar{y}} T_{x_2 x_2^\prime \bar{y} y^\prime}} \label{eq:edgeM} \\
    \tilde{T}^E_{x x^\prime y y^\prime} =&& \sum_{\bar{X},\bar{X}^\prime=1}^{D^2}{
        M^E_{\bar{X}\bar{X}^\prime yy^\prime}
        \tilde{O}_{\bar{X},x}
        \tilde{O}_{\bar{X}^\prime,x^\prime}
    }
\end{eqnarray}
where we use the same orthogonal matrices $\tilde{O}_{X,x}$ that we use to coarse grain $T$.

Note that the edge tensors do not impact the average value of observables measured on the space of closed paths, only the building of the worms. However, selecting correct edge tensors is necessary to optimize the dynamics of the worms, which impacts the convergence rate of all observables.  In addition, the edge tensors are essential for the measurement of the two-point function, which is measured using the open worm states \cite{Prokofev2001WormAlgorithms}.  Note that in Eq.~\eqref{eq:edgeM} we chose to put the edge tensor on a specific site of the coarse-grained supersite.  
Hence, the two-point function computed in the coarse-grained model is ``decimated'' to specific sites on the original lattice. For instance, consider the two-point functions after zero and one coarse graining steps, $c_{\mathrm{RG0}}(\mathbf{r})$ and $c_{\mathrm{RG1}}(\mathbf{r})$. Up to errors due to the truncation, we expect to find $c_{\mathrm{RG1}}(\mathbf{r})= c_{\mathrm{RG0}}(2\mathbf{r})$, {\em i.e.} $c_{\mathrm{RG1}}(\mathbf{r})$ gives the two-point function on even sites of the original lattice. We call this a decimated two-point function.  Note that at the critical point, correlation functions are expected to be power-law decaying, except for lattice effects at short distances.  Decimation does not affect the power laws, but it does reduce the short-distance lattice effects, giving a better approximation to the continuum limit.

For more details about the Tensor Worm Monte Carlo and edge tensors, see App.~\ref{app:MCmethods}.

\section{\label{sec:Discussion} Discussion}

In this work, we introduced a method for real space coarse-graining that allowed us to derive an RG-improved action for use in Monte Carlo simulations. To demonstrate the method, we applied it to the XY model on the cubic lattice and used it to compute critical exponents and universal amplitudes.  We found agreement with published values, showing that our method does not affect the universality class of the transition.  We found that coarse-graining significantly reduced the leading correction to scaling, allowing us to obtain critical properties with high accuracy on relatively small system sizes. For some observables, the second correction to scaling was reduced as well.  

In addition, we computed the two-point function at criticality.  We showed that we can obtain the two-point function of a large lattice by coarse-graining the model and measuring the two-point function on a smaller lattice. In addition, we found that the coarse-grained two-point function restores the emergent spherical symmetry at the critical point.  This reduced sensitivity to the underlying lattice suggests that the improved action approximates closely the continuum field theory. 

These successes highlight the power of the method. RG-improved actions may be useful in other applications.  A case in point is the extraction of dynamical properties near quantum critical points \cite{GPA,GPAA}, which presents a challenge for a number of reasons: To obtain real-time dynamical correlation functions from Monte Carlo simulations, it is necessary to perform an analytical continuation of the Euclidean time data, a procedure that requires numerical data of the highest quality.  This is difficult near the critical point, where the divergent correlation length means that large system sizes must be studied to perform finite-size scaling analyses. These efforts are further hindered by the presence of non-universal contributions at short distances, which generate corrections to scaling that can be sizable, making it difficult to discern universal properties from the large non-universal backgrounds \cite{lucas2017dynamical}.  

We expect our method to alleviate these problems.  In addition to the reduction in non-universal properties, the enhanced spherical symmetry can play an important role in analytical continuation.  Typically, when analytical continuation is done, only the data on the principal axes, such as $(k_x,0,0)$, is used.  However, the emergent spherical symmetry may allow the use of all long wave-length points, regardless of their direction relative to the lattice, resulting in a very significant increase in statistics.

As another potential application, our method may allow one to distinguish between weakly-first-order and continuous transitions in systems where there is controversy regarding the nature of the transition. This typically requires extremely large systems to resolve.  By performing several RG steps prior to the Monte Carlo simulations, it may be possible to reduce significantly the system sizes needed to settle such questions.

Further refinement of our method is possible. At first glance, coarse-graining reduces all corrections to scaling, suppressing subleading corrections more rapidly than leading corrections.  However, this is not what we observed, due to truncation.  The truncation of tensor models is known to introduce perturbations (relevant and irrelevant) at the critical point.  The relevant perturbations cause the appearance of a finite correlation length, which is equivalent to a shift in the critical point \cite{Ueda2017CriticalBehaviorIcosahedron,Tagliacozzo2008ScalingEntanglementMPS,Ueda2020FiniteScalingAnalysisBKT,Pirvu2012MatrixProductStateFiniteSize,Ueda2014DoublingEntanglementSpectrum,Ueda2023Finite-sizeBondDimensionEffectsTNR,Pollmann2009TheoryFiniteEntanglementScaling,Calabrese2008Entanglementspectrum}. The irrelevant perturbations, as we see in our analysis, add corrections to scaling to the model, thus limiting our method's ability to suppress them.

Increasing the bond dimension is expected to alleviate this issue.   In our work, we truncated down to $11$ states. This limit can be pushed further. For instance, in Ref.~\onlinecite{Xie2012RGbyHOSVD}, HOTRG done with $16$ states is reported. Alternatively, using different TRG methods such as ATRG and triad TRG \cite{adachi2020ATRG,kadoh2019TriagTRG,bloch2021TRGO2model} could potentially allow the use of more states and provide an improvement. 

\begin{acknowledgments}

This work was supported by the Israel Science Foundation under grant No.~2005/23.
    
\end{acknowledgments}

\appendix

\section{\label{app:NNHOTRG} Positive tensor truncation (PTT)}

In this appendix, we provide some of the details of the PTT described briefly in Sec.~\ref{subsec:generalMethod}. We write down the partitioning of the states into corner states explicitly and explain the motivation for this partitioning in more detail.

\subsection{\label{subsec:XYTruncation} Truncation of the XY Model}

As mentioned in Sec.~\ref{subsec:generalMethod}, we truncate the bond currents of the XY model by keeping states with $N_b^1,N_b^2 \leq 2$. This brings the number of bond states down to $9$ states, enumerated in Table~\ref{tab:XYstates}.%
\begin{table}
    \caption{\label{tab:XYstates} States kept in the truncated XY model.}
    \begin{ruledtabular}
    \begin{tabular}{c|ccc}
         \# & $N_b^1$ & $N_b^2$ & Total current \\ \hline
        1 & 0 & 0 & 0  \\
        2 & 1 & 1 & 0  \\
        3 & 1 & 0 & 1  \\
        4 & 0 & 1 & -1 \\
        5 & 2 & 1 & 1  \\
        6 & 1 & 2 & -1 \\
        7 & 2 & 2 & 0  \\
        8 & 2 & 0 & 2  \\
        9 & 0 & 2 & -2 
    \end{tabular}
    \end{ruledtabular}
\end{table}
We will label states by these numbers in what follows.

\subsection{\label{HOTRGStates} States of the HOSVD}

To understand the motivation behind the corner states used in the PTT, we examine the states of the HOSVD after one coarse-graining step.

We perform a single full coarse graining step (on all three dimensions) using the HOSVD truncation at the critical point, combining $2^3=16$ sites into one supersite. The interaction with neighboring supersites is through $4$ bonds which form the coarse-grained superbond.  States have a well-defined net current, $I$, defined as the difference between outgoing and ingoing currents through the superbond. The following states, depicted in Table~\ref{tab:HOTRGstates}, have the largest spectral weight (in descending order)
\begin{enumerate}
    \item A symmetric state with zero net current, $I=0$.
    \item Two symmetric states with $I=\pm 1$.
    \item Four anti-symmetric states, in the left-right and up-down directions,  with $I=\pm 1$.
    \item Two symmetric states with $I=\pm 2$.
    \item Two anti-symmetric states with a checkered pattern with $I=\pm 1$.
\end{enumerate}

\begin{table}
    \caption{\label{tab:HOTRGstates} First 11 kinds of states of the HOSVD up to $I=1$, at the critical point, after one coarse graining step.}
    \begin{ruledtabular}
    \begin{tabular}{c|c}
         Example &  \makecell{Number of \\ states with \\ similar symmetry }\\ \hline
         $
         \begin{array}{|c|c|} \hline
             0 & 0 \\ \hline
             0 & 0 \\ \hline
         \end{array}
         $
          & 1 \\ \hline
          
          $
         \begin{array}{|c|c|} \hline
             +1 & 0 \\ \hline
             0 & 0 \\ \hline
         \end{array}
         +
         \begin{array}{|c|c|} \hline
             0 & 0 \\ \hline
             +1 & 0 \\ \hline
         \end{array}
         +
         \begin{array}{|c|c|} \hline
             0 & +1 \\ \hline
             0 & 0 \\ \hline
         \end{array}
         +
         \begin{array}{|c|c|} \hline
             0 & 0 \\ \hline
             0 & +1 \\ \hline
         \end{array}
         $
          & 2 \\ \hline

            $
         \begin{array}{|c|c|} \hline
             +1 & 0 \\ \hline
             0 & 0 \\ \hline
         \end{array}
         +
         \begin{array}{|c|c|} \hline
             0 & 0 \\ \hline
             +1 & 0 \\ \hline
         \end{array}
         -
         \begin{array}{|c|c|} \hline
             0 & +1 \\ \hline
             0 & 0 \\ \hline
         \end{array}
         -
         \begin{array}{|c|c|} \hline
             0 & 0 \\ \hline
             0 & +1 \\ \hline
         \end{array}
         $
          & 4 \\ \hline

        \makecell{
            $
         \begin{array}{|c|c|} \hline
             +1 & 0 \\ \hline
             +1 & 0 \\ \hline
         \end{array}
         +
         \begin{array}{|c|c|} \hline
             0 & 0 \\ \hline
             +1 & +1 \\ \hline
         \end{array}
         +
         \begin{array}{|c|c|} \hline
             0 & +1 \\ \hline
             0 & +1 \\ \hline
         \end{array}
         +
         $\\
         $+
         \begin{array}{|c|c|} \hline
             +1 & +1 \\ \hline
             0 & 0 \\ \hline
         \end{array}
         +
         \begin{array}{|c|c|} \hline
             +1 & 0 \\ \hline
             0 & +1 \\ \hline
         \end{array}
         +
         \begin{array}{|c|c|} \hline
             0 & +1 \\ \hline
             +1 & 0 \\ \hline
         \end{array}
         $}
          & 2 \\ \hline
        $
         \begin{array}{|c|c|} \hline
             +1 & 0 \\ \hline
             0 & 0 \\ \hline
         \end{array}
         +
         \begin{array}{|c|c|} \hline
             0 & 0 \\ \hline
             0 & +1 \\ \hline
         \end{array}
         -
         \begin{array}{|c|c|} \hline
             0 & 0 \\ \hline
             +1 & 0 \\ \hline
         \end{array}
         -
         \begin{array}{|c|c|} \hline
             0 & +1 \\ \hline
             0 & 0 \\ \hline
         \end{array}
         $
          & 2 \\ \hline
    \end{tabular}
    \end{ruledtabular}
\end{table}
The states with negative values, which promote negative values in the renormalized tensor $\tilde{T}$, are the anti-symmetric states.  The $I=1$ states can be written as linear combinations of corner states which are positive.  Hence, by using corner states in the PTT, we preserve most of the spectral weight in the HOSVD while maintaining non-negativity.

\subsection{\label{subsec:StateDist} State Distribution in the PTT}

As mentioned in Sec.~\ref{subsec:generalMethod}, the PTT requires us to partition the states for the truncation. In this section we explicitly describe the partitioning.

\subsubsection{The First Step}

On the first coarse-graining step we choose the state partitioning used to construct the orthogonal matrix $O_{X,x}$ according to Table \ref{tab:StateDistriubtion1}.%
\begin{table}
    \caption{\label{tab:StateDistriubtion1} State partitioning on the first step.}
    \begin{ruledtabular}
    \begin{tabular}{c|*{9}{c}}
    $x_{1}\backslash x_{2}$
          & 1  & 2  & 3  & 4  & 5  & 6  & 7 & 8 & 9   \\ \hline
        1 & 1  & 1  & 4  & 8  & 4  & 8  & 1 & 10 & 11 \\
        2 & 1  & 1  & 4  & 8  & 4  & 8  & 1 & 10 & 11 \\
        3 & 2  & 2  & 10 & 1  & 10 & 1  & 2 &    & 9  \\
        4 & 6  & 6  & 1  & 11 & 1  & 11 & 6 & 5  &    \\
        5 & 2  & 2  & 10 & 1  & 10 & 1  & 2 &    & 9  \\
        6 & 6  & 6  & 1  & 11 & 1  & 11 & 6 & 5  &    \\
        7 & 1  & 1  & 4  & 8  & 4  & 8  & 1 & 10 & 11 \\
        8 & 10 & 10 &    & 3  &    & 3  & 10 &   & 1 \\
        9 & 11 & 11 & 7  &    & 7  &    & 11 & 1 & 
    \end{tabular}
    \end{ruledtabular}
\end{table}
It is to be read as follows: the rows designate the number of state $x_1$, the columns designate the number of the state $x_2$, and the value at row $x_1$ and column $x_2$ is the $x$ for which $O_{x_1x_2,x}$ is non-zero. For instance, $x=1$ is a superstate with  $I=0$, hence it includes all pairs $X=x_1x_2$ that are neutral. We include four superstates with $I=1$ ($x=2,3,4,5$) that are distinguished according to how the currents are distributed in space and, similarly, four superstates with $I=-1$ ($x=6,7,8,9$). Finally, we keep two superstates ($x=10,11$) with $I=\pm 2$.  Note that some entries in the table are missing.  These are combinations of states with total current $|I|>2$, which are truncated.

To determine the new states in the $x$ direction on the first step of the RG (in 3D), we compute $A=M^\prime M^{\prime\dagger}$, where $M^\prime$ was defined in Eq. \eqref{eq:Mprime}.  For each superstate $x$, truncate $A$ to $A_x$ in the relevant subsector of states $x_1,x_2$, according to Table \ref{tab:StateDistriubtion1}. Then, find the first eigenvector (with the largest eigenvalue) of $A_x$ in the subsector. This eigenvector defines the values of $O_{X,x}$, and it can always be chosen to be non-negative. For example, for $x=1$ we obtain:
\begin{eqnarray}
    &&O_{x_1x_2,1} = \nonumber\\
    &&\begin{pmatrix}
        O_{11} & O_{12} & 0        & 0        & 0        & 0        & O_{17} & 0        & 0        \\
        O_{21} & O_{22} & 0        & 0        & 0        & 0        & O_{27} & 0        & 0        \\
        0        & 0        & 0        & O_{34} & 0        & O_{36} & 0        & 0        & 0        \\
        0        & 0        & O_{43} & 0        & O_{45} & 0        & 0        & 0        & 0        \\
        0        & 0        & 0        & O_{54} & 0        & O_{56} & 0        & 0        & 0        \\
        0        & 0        & O_{63} & 0        & O_{65} & 0        & 0        & 0        & 0        \\
        O_{71} & O_{72} & 0        & 0        & 0        & 0        & O_{77} & 0        & 0        \\
        0        & 0        & 0        & 0        & 0        & 0        & 0        & 0        & O_{89} \\
        0        & 0        & 0        & 0        & 0        & 0        & 0        & O_{98} & 0
    \end{pmatrix}
\end{eqnarray}
where the specific non-zero values are chosen according to the SVD.

\subsection{Following Steps}
The state distribution used from the second step onward is shown in Table \ref{tab:StateDistribution2}.
\begin{table}
    \caption{\label{tab:StateDistribution2} State distribution used for steps after the first step.}
    \begin{ruledtabular}
    \begin{tabular}{c|*{11}c}
        $x_{1}\backslash x_{2}$
          & 1  & 2  & 3  & 4  & 5  & 6  & 7  & 8  & 9  & 10 & 11   \\ \hline
        1 & 1  & 4  & 4  & 5  & 5  & 8  & 8  & 9  & 9  & 10 & 11 \\
        2 & 2  & 10 & 10 & 10 & 10 & 1  & 1  & 1  & 1  &    & 9 \\
        3 & 2  & 10 & 10 & 10 & 10 & 1  & 1  & 1  & 1  &    & 9 \\
        4 & 3  & 10 & 10 & 10 & 10 & 1  & 1  & 1  & 1  &    & 8 \\
        5 & 3  & 10 & 10 & 10 & 10 & 1  & 1  & 1  & 1  &    & 8 \\
        6 & 6  & 1  & 1  & 1  & 1  & 11 & 11 & 11 & 11 & 5  &   \\
        7 & 6  & 1  & 1  & 1  & 1  & 11 & 11 & 11 & 11 & 5  &   \\
        8 & 7  & 1  & 1  & 1  & 1  & 11 & 11 & 11 & 11 & 4  &   \\
        9 & 7  & 1  & 1  & 1  & 1  & 11 & 11 & 11 & 11 & 4  &   \\
       10 & 10 &    &    &    &    & 3  & 3  & 2  & 2  &    & 1 \\
       11 & 11 & 7  & 7  & 6  & 6  &    &    &    &    & 1  &   \\
    \end{tabular}
    \end{ruledtabular}
\end{table}
As before, when constructing this table we chose to keep $11$ superstates.  The superstate $1$ is neutral, $I=0$.  The superstates $2,3,4,5$ are corner states with $I=1$, and $6,7,8,9$ are corner states with $I=-1$.  The superstates $10,11$ have $I=\pm 2$.

\subsection{Current Count and Distribution}

One may be worried about the suitability of the use of corner states and the sensibility of truncating states with net current larger than two when we perform subsequent RG steps, since as the face of the supersite grows, larger net currents are expected to pass through the face.

To check on this matter, we performed a simulation of the RG0 model and measured the net current going through the faces of cubes of various sizes. Table~\ref{tab:currentCount}%
\begin{table}
    \caption{\label{tab:currentCount} Fraction of faces with a certain absolute value of net current, in percent ($\%$).  Computed using RG0 model on a $64\times 64 \times 64 $ lattice and measured the net current going through the faces of cubes of various sizes $l\times l \times l$.}
    \begin{ruledtabular}
    \begin{tabular}{c|ccccc}
        $l$ & $I=0$ & $|I|=1$ & $|I|=2$ & $|I|=3$ & $|I|=4$ \\ \hline
        $1$ & $94.7$ & $5.3$ & $0$ & $0$ & $0$ \\
        $2$ & $86.6$ & $13.0$ & $0.4$ & $0$ & $0$ \\
        $4$ & $72.2$ & $25.6$ & $2.1$ & $0.1$ & $0$ \\
        $8$ & $52.4$ & $39.2$ & $7.4$ & $0.9$ & $0.1$ \\
        $16$& $33.9$ & $44.0$ & $16.7$ & $4.4$ & $0.9$ \\
        $32$& $21.9$ & $37.2$ & $23.3$ & $11.3$ & $4.4$
    \end{tabular}
    \end{ruledtabular}
\end{table}
shows, for each cube size $l$, the percentage of encountered faces with a specific net current (absolute value).

We see that for $l=1$ $95\%$ of the faces have zero net current, and about $5\%$ have $|I|=1$. For $l=2$, only $0.4\%$ of the faces have $|I|=2$. For $l=8$, only $7.4\%$ of the faces have $|I|=2$, and less than $1\%$ have $|I|=3$. For $l=32$, however, already $23\%$ of the faces have $|I|=2$, and $11\%$ of the faces have $|I|=3$. This is an indication that our truncation is reasonable for RG3, which corresponds to $l=8$, but that it becomes less faithful as we continue to RG5. This may explain why the corrections to scaling stop improving as we coarse grain further, since states with higher net current are necessary to accurately capture the current fluctuations for large $l$.

\begin{figure}
    \includegraphics{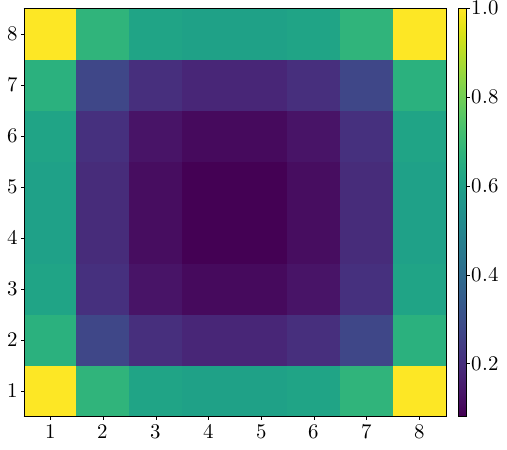}
    \caption{\label{fig:CurrentDist} Current distribution on faces of size $8\times 8$ with total current $I=1$ in a simulation of model RG0 on lattice size $64\times64\times64$. Normalized so that the maximum is 1.}
\end{figure}

Another interesting picture is the current distribution on the face of a cube.  Figure~\ref{fig:CurrentDist}%
shows the current distribution for $8\times 8$-sized faces, averaged over configurations with net current $I=1$. We see that the current is mostly localized on the edges and especially on the corners of the face. This occurs since most of the closed worms are small loops.  Hence, they only produce a net current when they pass near the edges or corners of the face. This further supports the use of corner states.

\subsection{\label{subsec:Normalization} Normalization}

After each step, we normalize the tensors by dividing by the largest element in the tensor:
\begin{equation}
    \Tilde{T}_{xx^\prime yy^\prime zz^\prime} \to
    \frac{\Tilde{T}_{xx^\prime yy^\prime zz^\prime}}
    {\max \Tilde{T}_{xx^\prime yy^\prime zz^\prime}}
\end{equation}
Without the normalization, the values of the tensors grow after each coarse-graining step, which hinders numerics. Since Monte Carlo simulations use only the ratios of the tensors, the simulations are unaffected by the normalization.

\subsection{\label{subsec:Symmetrization} Symmetrization}
Since we perform the coarse graining one axis at a time, the resulting tensor is anisotropic. However, since we expect isotropic results, after each full RG step we perform a {\em symmetrization} of the coarse-grained tensor.

The symmetrization is done by averaging over all the possible permutations of the axes:
\begin{eqnarray}
    T_{\text{sym}} = && T
    + T(xyz\to yxz) \nonumber\\
    && + T(xyz\to zyx) 
    + T(xyz\to xzy) \nonumber\\
    && + T(xyz\to zxy)
    + T(xyz\to yzx)
\end{eqnarray}

\begin{figure}
    \includegraphics{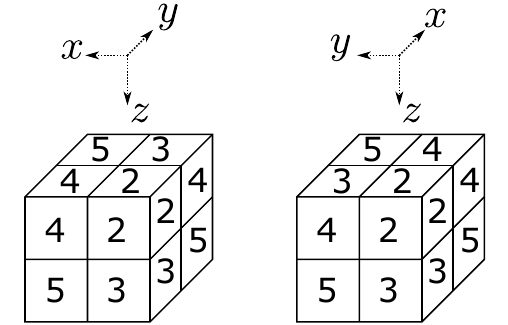}
    \caption{\label{fig:TensorSymmetrization} Permutation of $x \leftrightarrow y$ swaps the corner states $3$ and $4$ on the $z$ face.}
\end{figure}

It is important to note that the axis permutations do not amount only to index swaps, but they also alter the states themselves, since the states are not invariant under this kind of transformation. An example is shown in Fig.~\ref{fig:TensorSymmetrization},%
where an $x \leftrightarrow y$ transformation also requires to swap states the corner states $3$ and $4$ (and their corresponding states with negative currents) on the $z$ direction.

The symmetrization is easy to carry out because of the predetermined corner states we use in the PTT. Since we know the states we know how they transform under axis permutations.

\section{\label{app:MCmethods} Monte Carlo Methods}

Our Monte Carlo simulations use a version of the worm algorithm of Ref.~\onlinecite{Prokofev2001WormAlgorithms} with the tensor network action. In this Appendix, we describe how we perform importance sampling on tensor networks based on the worm algorithm.  First, we describe the standard worm algorithm \cite{Prokofev2001WormAlgorithms}, and then describe the tensor worm algorithm that we use.

\subsection{\label{subsec:WormAlgorithm} Worm Algorithm}

The Worm algorithm is based on the high-temperature expansion of the action, which is used to rewrite the action as a sum over closed paths, as mentioned in Section \ref{subsec:TensorRepresentation}. The worm algorithm samples the closed path configurations by enlarging the state space to include states with a single open worm. When the worm closes, the system returns to the original state space.

The extension to open worms is achieved by inserting a source and a drain at sites $i_1,i_2$ of the lattice, respectively. The corresponding weight is:
\begin{equation}\label{eq:gXY}
     w(i_1-i_2) = \int \left(\prod_i{\operatorname{d}\psi_i\operatorname{d}\psi_i^*}\right)
    {\psi_{i_1} \psi_{i_2}^* e^{-H_{O(2)}(\{\psi_i\})}}
\end{equation}
which, after a high-temperature expansion, yields (in the $\lambda\to\infty$ limit of the hard-spin XY model):
\begin{equation}\label{eq:gworm}
    w(i_1-i_2) = \sum_{CP^*}{
    \left ( 
        \prod_b{\frac{g^{N_b^1+N_b^2}}{N_1!N_2!}}
    \right )
    }\,.
\end{equation}
This is similar to Eq.~\eqref{eq:Zworm}, except that the sum is over all directed closed paths with one open path connecting sites $i_1$ and $i_2$.

These weights can be used to define a Monte Carlo algorithm.
The states are determined by the points $i_1,i_2$ and the bond numbers $N_b^1,N_b^2$ on every bond $b$ on the lattice.
The worm updates are as follows:
\begin{itemize}
    \item If $i_1=i_2$ perform a ``move'' step with probability $p$ or a ``shift'' step with probability $1-p$.  If $i_1\neq i_2$ always perform a ``shift''.
    \item A ``move'' step moves both $i_1$ and $i_2$ to a new random lattice site. The acceptance ratio for such a step is $1$.
    \item A ``shift'' step shifts $i_1$ to a random neighboring site $j$ along a bond $b$. It also selects at random whether to add or remove current when moving from site $i_1$ to site $j$. Adding a current increases the bond number $N_b^1$ by $1$, indicating more current flowing from $i_1$ to $j$. Removing a current decreases the bond number $N_b^2$ by $1$, indicating less current flowing from $j$ to $i_1$. In both cases, $i_1$ is shifted to $j$. The acceptance ratio for a ``shift'' step is
    \begin{equation}
        P_\text{sh}(i_1\to j;N_b^1\to N_b^1+1) = r
        \frac{g}{N_b^1+1}
    \end{equation}
    \begin{equation}
        P_\text{sh}(i_1\to j;N_b^2\to N_b^2-1) = r
        \frac{N_b^2}{g}
    \end{equation}
    where $r$ is
    \begin{equation}
        r = 
        \begin{cases}
            1/(1-p) & i_1=i_2\ \mathrm{(worm\ opens)} \\
            (1-p) & j=i_2 \  \ \mathrm{(worm\ closes)} \\
            1 & \text{otherwise}
        \end{cases}
    \end{equation}
\end{itemize}
When the worm closes, we return to the state space of the closed worm action. Then, we can calculate estimators.

\subsection{\label{subsec:TensorAlgoEdgeTensors} Tensor Worm Algorithm and Edge Tensors}

In Sec.~\ref{subsec:TensorRepresentation} we wrote the worm partition function in the language of tensor networks. Here we describe a Monte Carlo algorithm based on the tensor partition function.
For simplicity, we explain it only on the square lattice, but the algorithm is easily extended to higher dimensions.

In the tensor language the worm partition function is written as a tensor network:
\begin{equation}
    Z = \sum_{\{x_i,y_i\}}\prod_i{T_{x_ix_i^\prime y_iy_i^\prime}}
\end{equation}
where the tensors impose current conservation. To perform a worm algorithm, we also extend the tensor description to the open worm weights of Eq. \eqref{eq:gworm}
\begin{eqnarray}\label{eq:gtensoropen}
    w(i_1-i_2;i_1\neq i_2) = &&\sum_{\{x_i,y_i\}}
    T^{E1}_{x_{i_1}x_{i_1}^\prime y_{i_1}y_{i_1}^\prime}
    T^{E1}_{x_{i_2}x_{i_2}^\prime y_{i_2}y_{i_2}^\prime} \times \nonumber\\
    && \times \prod_{i\notin \{i_1,i_2\}}{T_{x_ix_i^\prime y_iy_i^\prime}}
\end{eqnarray}
\begin{equation}\label{eq:gtensorclosed}
    w(i_1-i_2;i_1=i_2) = \sum_{\{x_i,y_i\}}
    T^{E0}_{x_{i_1}x_{i_1}^\prime y_{i_1}y_{i_1}^\prime}
    \prod_{i\neq i_1}{T_{x_ix_i^\prime y_iy_i^\prime}}
\end{equation}
where $T^{E0}_{x_{i}x_{i}^\prime y_{i}y_{i}^\prime},T^{E1}_{x_{i}x_{i}^\prime y_{i}y_{i}^\prime}$ are ``edge tensors'', describing the weights of the edges of the worm. $T^{E1}_{x_{i}x_{i}^\prime y_{i}y_{i}^\prime}$ are the weights when the drain and the source are on separate sites $i_1\neq i_2$, while $T^{E0}_{x_{i}x_{i}^\prime y_{i}y_{i}^\prime}$ are the weights when the drain and the source are on the same site $i_1=i_2$. For the XY model:
\begin{equation}\label{eq:edgetensorXY1}
    T^{E1}_{N_b^1,N_b^2} = \sqrt{\frac{g^{J_{\text{in}}+J_{\text{out}}}}
    {\displaystyle \prod_{b\in \mathrm{site}}N_{b}^{1}!N_{b}^{2}!}}
    \ \delta_{1,\big| J_{\text{in}} - J_{\text{out}} \big|}
\end{equation}
\begin{equation}\label{eq:edgetensorXY2}
    T^{E0}_{N_b^1,N_b^2} = \sqrt{\frac{g^{J_{\text{in}}+J_{\text{out}}}}
    {\displaystyle \prod_{b\in \mathrm{site}}N_{b}^{1}!N_{b}^{2}!}}
     \ \delta_{J_{\text{in}},J_{\text{out}}}
\end{equation}
Since the two kinds of edge tensors are nonzero for different currents, in practice we keep track of them using a combined edge tensor:
\begin{equation}
    T^E_{x_{i}x_{i}^\prime y_{i}y_{i}^\prime} = T^{E0}_{x_{i}x_{i}^\prime y_{i}y_{i}^\prime} + T^{E1}_{x_{i}x_{i}^\prime y_{i}y_{i}^\prime}
\end{equation}
where for any set of indices ${x_{i}x_{i}^\prime y_{i}y_{i}^\prime}$ at most one of the terms is non-zero.

The steps of the tensor worm algorithm are similar to the steps of the standard worm algorithm described in the previous section, except that the acceptance ratios are set by the tensors:
\begin{itemize}
    \item If $i_1=i_2$ suggest a ``move'' with probability $p$ or suggest a ``shift'' with probability $1-p$. If $i_1\neq i_2$, always suggest a ``shift''.
    \item A ``move'' moves both $i_1$ and $i_2$ to another random point on the lattice. The acceptance ratio for such a step is $1$.
    \item A ``shift'' shifts $i_1$ to a random neighboring site. The shift always increases the net current on the bond between the sites, changing the current from $J$ to $J+1$. A random bond state with net current $J+1$ is chosen. For instance, the acceptance ratio for a ``shift'' in the positive $x$ direction (on the bond $x_i^\prime$) with a new bond state ${\xi_i^\prime}$ is
    \begin{equation}
        P_\text{sh}(i\to j;i,j\neq i_2) = s
        \frac{T_{x_{i}\xi_{i}^\prime y_{i}y_{i}^\prime}}
        {T^E_{x_{i}x_{i}^\prime y_{i}y_{i}^\prime}}
        \frac{T^E_{\xi_{i}^\prime x_{j}^\prime y_{j}y_{j}^\prime}}
        {T_{x_{i}^\prime x_{j}^\prime y_{j}y_{j}^\prime}}
    \end{equation}
    where $s$ is the ratio between the number of states with current $J+1$ and states with current $J$.
    
    For the opening of a worm (if $i_1=i_2$), the probability is:
    \begin{equation}
        P_\text{sh}(i=i_2\to j) = s
        \frac{1}{1-p}
        \frac{T^E_{x_{i}\xi_{i}^\prime y_{i}y_{i}^\prime}}
        {T^E_{x_{i}x_{i}^\prime y_{i}y_{i}^\prime}}
        \frac{T^E_{\xi_{i}^\prime x_{j}^\prime y_{j}y_{j}^\prime}}
        {T_{x_{i}^\prime x_{j}^\prime y_{j}y_{j}^\prime}}
    \end{equation}
    For the closing of a worm (if $j=i_2$), the probability is:
    \begin{equation}
        P_\text{sh}(i\to j=i_2) = s
        (1-p)
        \frac{T_{x_{i}\xi_{i}^\prime y_{i}y_{i}^\prime}}
        {T^E_{x_{i}x_{i}^\prime y_{i}y_{i}^\prime}}
        \frac{T^E_{\xi_{i}^\prime x_{j}^\prime y_{j}y_{j}^\prime}}
        {T^E_{x_{i}^\prime x_{j}^\prime y_{j}y_{j}^\prime}}
    \end{equation}
\end{itemize}

As in the standard worm algorithm, we calculate estimators when the worm closes and we return to the closed-worms state space.

\subsubsection{Time Comparison with Standard Worm Algorithm}

In this section, we compare the run times of the tensor worm algorithm and the standard worm algorithm.
We use the results of several Monte Carlo simulations on a $14 \times 14 \times 14$ lattice of each model at the critical point. For each model, we calculated the average run time $t_\mathrm{run}$ of the Monte Carlo and the average statistical errors of $R_x$, $\delta R_x$. We also calculated the relative time penalty factor, defined as $s=\frac{t_\mathrm{run}}{t_\mathrm{run}^0}(\frac{\delta R_x}{\delta R^0_x})^2$ where $t_\mathrm{run}^0$ and $\delta R^0_x$ are the run time and error in $R_x$ for the standard worm algorithm. The relative time factor takes into account the errors in observables, since errors in Monte Carlo simulations decrease as $\frac{1}{\sqrt{t_\mathrm{run}}}$. Hence, $s$ is the penalty factor needed to obtain results with the same statistical errors as the standard worm algorithm. The results are in Table \ref{tab:runtimeComparison}.%
\begin{table}
    \caption{\label{tab:runtimeComparison} Run time comparison between the standard worm algorithm and the tensor worm algorithm for various RG stages at the critical point. The values of the time and errors are arbitrary, only relative values are meaningful.}
    \begin{ruledtabular}
    \begin{tabular}{c|ccc}
        Algorithm & Time (hours) & $\delta R_x$ & \makecell{Relative Time \\ Penalty} \\ \hline
        Standard 
        & $6.5$        & $0.00018$    & 1.0 \\
        Tensor RG0 
        & $13.0$        & $0.00022$    & 2.8 \\
        Tensor RG1 
        & $13.6$        & $0.00018$    & 2.0 \\
        Tensor RG3 
        & $11.1$        & $0.00026$    & 3.6 \\
        Tensor RG5 
        & $14.0$        & $0.00028$    & 5.1 \\
    \end{tabular}
    \end{ruledtabular}
\end{table}

The time factor of RG0 represents the time penalty of the worm algorithm compared to the standard worm algorithm, before any coarse graining. By profiling the code, we found that the primary source of the penalty is the lookup time of the weights in the tensor $T$ during the worm steps.  For other RG steps, some of the time penalty may be due to the modification of the dynamics by the coarse graining, which can affect the statistical errors of observables at the  
critical point.  It may be possible to reduce the time penalty by a different choice of edge tensors, which could be optimized to maximize the efficiency of the method without biasing observables that are measured in the state-space of closed worms.

\subsection{Observables in the Coarse-Grained Tensor Worm Algorithm}

Measuring observables on the tensor worm algorithm is different compared to the standard worm algorithm, especially after performing RG steps.

\subsubsection{Scalar Observables}

 Scalar observables are the easiest to measure. Similar to the case of the standard XY model, we keep track of the winding of currents around the lattice. Using the winding numbers we can calculate these observables. The calculation is the same for the tensor algorithm, since current remains a good quantum number of the tensor states.

\subsubsection{\label{subsec:2ptfuncmeasurement} Two-Point Correlation Function}

We can measure the two-point correlation function
\begin{equation}
    c(\textbf{r}) = \langle \psi_i \psi_{i+\textbf{r}} \rangle
\end{equation}
by using the open worm weight $w(\textbf{r})$ of Eq.~\eqref{eq:gworm} for the standard worm algorithm, and Eqs.~\eqref{eq:gtensoropen} and \eqref{eq:gtensorclosed} for the tensor algorithm. Thus, the estimator is simply $\delta_{i_2-i_1,\textbf{r}}$ \cite{Prokofev2001WormAlgorithms}.

Note that for the coarse-grained model we decimate the edge to a specific site, so after one RG step the two-point function corresponds to:
\begin{equation}\label{eq:decimation}
    c_{\text{RG1}}(\textbf{r}) = c_{\text{RG0}}(2\textbf{r})
\end{equation}
and similarly for more RG steps. We find that the coarse grained two-point function is aliased. Note that the functions $c_{\text{RG1}}(\textbf{r})$ and $c_{\text{RG0}}(\textbf{r})$ above are on the same effective lattice size, aka $L_{\text{RG0}}=2L_{\text{RG1}}$.

In momentum space the aliasing ``folds'' the correlation function in all three dimensions:
\begin{eqnarray} \label{eq:aliasing}
    \Tilde{c}^{(1)}&&(k_x,k_y,k_z) = 
    \Tilde{c}^{(0)}(\frac{k_x}{2},\frac{k_y}{2},\frac{k_z}{2}) 
     + \Tilde{c}^{(0)}(\frac{\pi - k_x}{2},\frac{k_y}{2},\frac{k_z}{2}) \nonumber \\
    && + \Tilde{c}^{(0)}(\frac{k_x}{2},\frac{\pi - k_y}{2},\frac{k_z}{2}) 
      +\Tilde{c}^{(0)}(\frac{k_x}{2},\frac{k_y}{2},\frac{\pi - k_z}{2}) \nonumber \\
    && +\Tilde{c}^{(0)}(\frac{\pi - k_x}{2},\frac{\pi -k_y}{2},\frac{k_z}{2}) 
     + \Tilde{c}^{(0)}(\frac{k_x}{2},\frac{\pi - k_y}{2},\frac{\pi -k_z}{2}) \nonumber \\
    && +\Tilde{c}^{(0)}(\frac{\pi -k_x}{2},\frac{k_y}{2},\frac{\pi - k_z}{2}) \nonumber \\
    && + \Tilde{c}^{(0)}(\frac{\pi-k_x}{2},\frac{\pi-k_y}{2},\frac{\pi-k_z}{2}) 
\end{eqnarray}

\section{Other Scalar Observables at Criticality}
\label{App:OtherObservables}

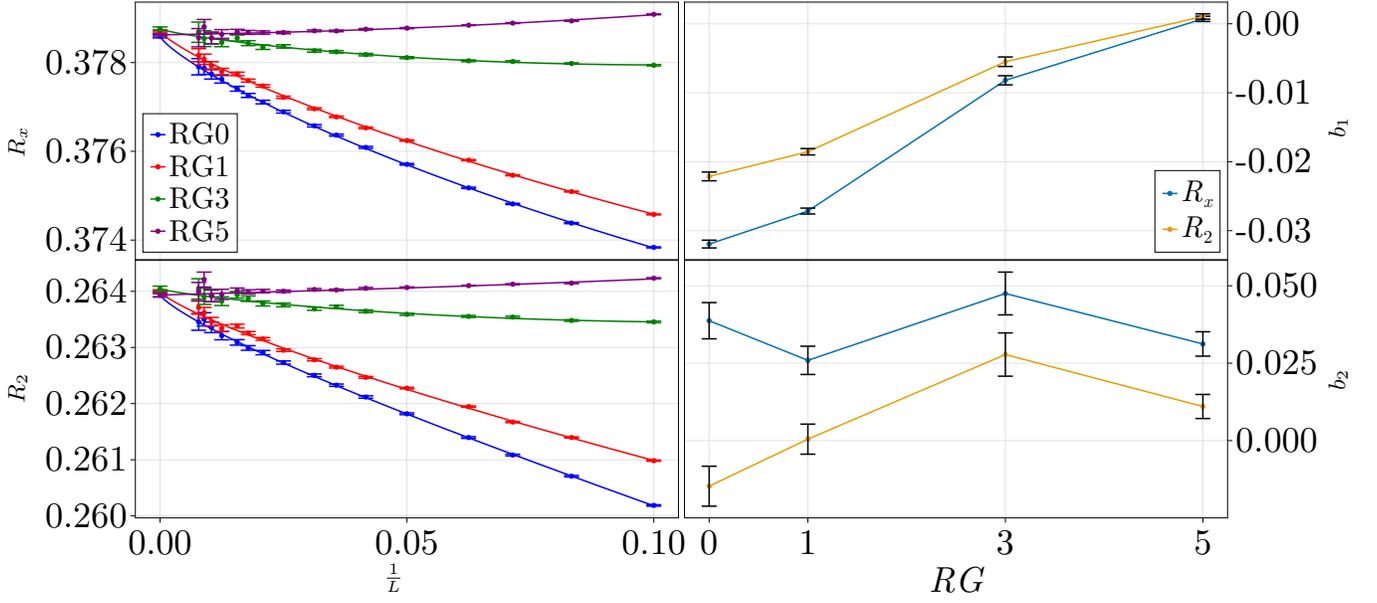
\begin{figure*}
    \begin{overpic}[width=.96\textwidth]{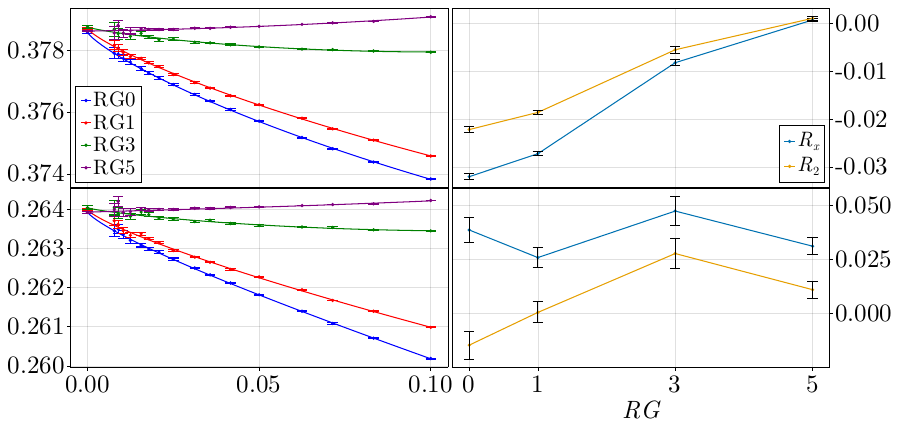}
    \put(-2, 35){\rotatebox{90}{$R_x$}}
    \put(-2, 16){\rotatebox{90}{$R_2$}}
    \put(100, 36.5){\rotatebox{90}{$b_1$}}
    \put(100, 17){\rotatebox{90}{$b_2$}}
    \put(27, 2){$\frac{1}{L}$}
    \end{overpic}
    \caption{\label{fig:DimensionlessObservablesCritEx} On the left panels, wrapping probabilities $R_x$ and $R_2$  as a function $\frac{1}{L}$ for different RG stages, evaluated at the critical point. Lines are fitted with $2$ corrections to scaling, according to Eq. \eqref{eq:ScalarFitForm}. The points at $\frac{1}{L}=0$ are the universal value $R^*$ according to the fits.  On the right panels, corrections to scaling from the values of $b_1$ and $b_2$ used in the fits.  The leading correction to scaling, $b_1$, is strongly suppressed for both observables.}
\end{figure*}

In this section we show results for the scalar observables $R_x$ and $R_2$ that are described in Sec.~\ref{subsec:ScalarObservables}. First, in Fig.~\ref{fig:DimensionlessObservablesCritEx}%
we show the scalar observables vs lattice size for each corase graining stage, similar to Fig.~\ref{fig:DimensionlessObservablesCrit}. Table~\ref{tab:correctionsOtherR} and Fig.~\ref{fig:DimensionlessObservablesCrit} shows the values of $R^*$ and the corrections to scaling amplitudes $b_1$ and $b_2$ found from the fits.

\begin{figure*}
    \begin{overpic}[width=.96\textwidth]{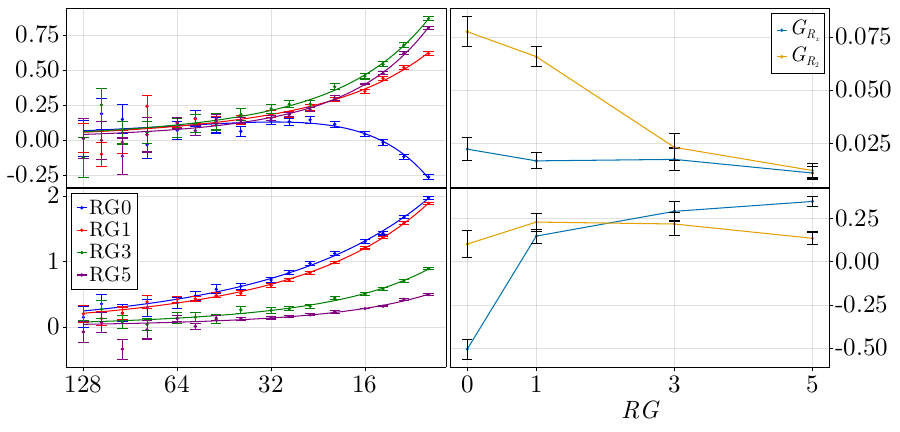}
    \put(-3, 28){\rotatebox{90}{$\log (G_{R_x}\, L^{-\frac{1}{\nu}}/R^*) \cdot 10^2$}}
    \put(-3, 7){\rotatebox{90}{$\log (G_{R_2} L^{-\frac{1}{\nu}}/R^*) \cdot 10^2$}}
    \put(100, 36){\rotatebox{90}{$b_1/R^*$}}
    \put(100, 16){\rotatebox{90}{$b_2/R^*$}}
    \put(28, 2){$L$}
    \end{overpic}
    \caption{\label{fig:DerivativeObservablesCritEx} On the left panels, normalized derivatives of $R_x$ and $R_2$ as a function of $L$ for different RG stages, evaluated at the critical point.  The derivatives have been rescaled by $L^{-1/\nu}$ with $\nu=0.67169(7)$ \cite{Hasenbusch2019MonteCarloClock}.  The plot is in log-log scale, such that a slope represents a correction to the cited critical exponent. Lines are fitted according to Eq.~\eqref{eq:DerivativeFittingForm}.  On the right panels, corrections to scaling from the values of $b_1/R^*$ and $b_2/R^*$ used in the fits.  The leading correction to scaling, $b_1/R^*$, is strongly suppressed for both observables.}
\end{figure*}
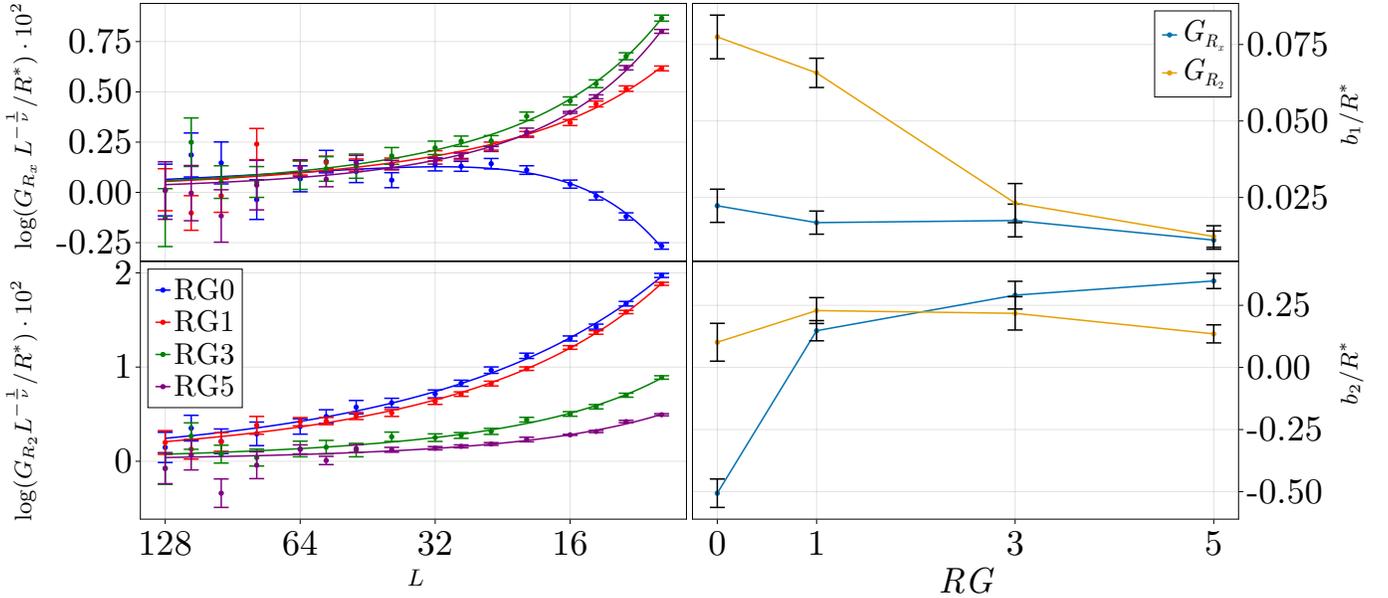

Next, we show the derivatives of these observables in Fig.~\ref{fig:DerivativeObservablesCritEx}%
, which is similar to Fig.~\ref{fig:DerivativeObservablesCrit} but for $G_{R_x}$ and $G_{R_2}$. Table~\ref{tab:correctionsOtherR} and Fig.~\ref{fig:DerivativeObservablesCrit} shows the corrections to scaling amplitudes.

\begin{table*}[t]
    \begin{ruledtabular}
    \begin{tabular}{c|lll|llr|rr|rr}
                 &  \multicolumn{3}{c|}{\,$R_x$} & \multicolumn{3}{c|}{$R_2$}  &  \multicolumn{2}{c|}{$G_{R_x}$} & \multicolumn{2}{c}{$G_{R_2}$}  \\
           & \multicolumn{1}{c}{$R^*$} & \multicolumn{1}{c}{$b_1$} & \multicolumn{1}{c|}{$b_2$} & \multicolumn{1}{c}{$R^*$} & \multicolumn{1}{c}{$b_1$} & \multicolumn{1}{c|}{$b_2$}
            & \multicolumn{1}{c}{$b_1/R^*$} & \multicolumn{1}{c|}{$b_2/R^*$} & \multicolumn{1}{c}{$b_1/R^*$} & \multicolumn{1}{c}{$b_2/R^*$}\\ \hline
        RG0   &
        $0.37859(4)$ & $-0.0319(6)$ & $0.039(6)$ & 
        $0.26394(4)$ & $-0.0221(6)$ & $-0.015(6)$ & 
        $0.022(5)$ & $-0.51(6)$ & 
        $0.077(7)$ & $0.10(8)$ \\
        RG1   & 
        $0.37871(3)$ & $-0.0272(4)$ & $0.026(5)$ & 
        $0.26400(3)$ & $-0.0186(5)$ & $0.000(5)$ & 
        $0.017(4)$ & $0.15(4)$ & 
        $0.066(5)$ & $0.23(5)$ \\
        RG3   & 
        $0.37876(4)$ & $-0.0082(7)$ & $0.048(7)$ & 
        $0.26405(5)$ & $-0.0055(7)$ & $0.028(7)$ & 
        $0.017(5)$ & $0.29(6)$ & 
        $0.023(6)$ & $0.22(7)$ \\
        RG5   & 
        $0.37862(3)$ & $\phantom{-}0.0007(4)$ & $0.031(4)$ & 
        $0.26393(3)$ & $\phantom{-}0.0011(4)$ & $0.011(4)$ & 
        $0.011(3)$ & $0.35(3)$ & 
        $0.012(4)$ & $0.14(4)$ \\
    \end{tabular}
    \end{ruledtabular}
    \caption{\label{tab:correctionsOtherR} Corrections to scaling.  Values of $R^*$, $b_1$ and $b_2$ used in the fits of the dimensionless observables to Eq.~\eqref{eq:ScalarFitForm} in Fig.~\ref{fig:DimensionlessObservablesCritEx}, and values of $b_1/R^*$ and $b_2/R^*$ used in the fits of the derivative observables to Eq.~\eqref{eq:DerivativeFittingForm} in Fig.~\ref{fig:DerivativeObservablesCritEx}.
    }
\end{table*}

\bibliography{apssamp}

\providecommand{\noopsort}[1]{}\providecommand{\singleletter}[1]{#1}%
\begin{thebibliography}{31}%
\makeatletter
\providecommand \@ifxundefined [1]{%
 \@ifx{#1\undefined}
}%
\providecommand \@ifnum [1]{%
 \ifnum #1\expandafter \@firstoftwo
 \else \expandafter \@secondoftwo
 \fi
}%
\providecommand \@ifx [1]{%
 \ifx #1\expandafter \@firstoftwo
 \else \expandafter \@secondoftwo
 \fi
}%
\providecommand \natexlab [1]{#1}%
\providecommand \enquote  [1]{``#1''}%
\providecommand \bibnamefont  [1]{#1}%
\providecommand \bibfnamefont [1]{#1}%
\providecommand \citenamefont [1]{#1}%
\providecommand \href@noop [0]{\@secondoftwo}%
\providecommand \href [0]{\begingroup \@sanitize@url \@href}%
\providecommand \@href[1]{\@@startlink{#1}\@@href}%
\providecommand \@@href[1]{\endgroup#1\@@endlink}%
\providecommand \@sanitize@url [0]{\catcode `\\12\catcode `\$12\catcode `\&12\catcode `\#12\catcode `\^12\catcode `\_12\catcode `\%12\relax}%
\providecommand \@@startlink[1]{}%
\providecommand \@@endlink[0]{}%
\providecommand \url  [0]{\begingroup\@sanitize@url \@url }%
\providecommand \@url [1]{\endgroup\@href {#1}{\urlprefix }}%
\providecommand \urlprefix  [0]{URL }%
\providecommand \Eprint [0]{\href }%
\providecommand \doibase [0]{https://doi.org/}%
\providecommand \selectlanguage [0]{\@gobble}%
\providecommand \bibinfo  [0]{\@secondoftwo}%
\providecommand \bibfield  [0]{\@secondoftwo}%
\providecommand \translation [1]{[#1]}%
\providecommand \BibitemOpen [0]{}%
\providecommand \bibitemStop [0]{}%
\providecommand \bibitemNoStop [0]{.\EOS\space}%
\providecommand \EOS [0]{\spacefactor3000\relax}%
\providecommand \BibitemShut  [1]{\csname bibitem#1\endcsname}%
\let\auto@bib@innerbib\@empty
\bibitem [{\citenamefont {Sachdev}(2011)}]{ssbook}%
  \BibitemOpen
  \bibfield  {author} {\bibinfo {author} {\bibfnamefont {S.}~\bibnamefont {Sachdev}},\ }\href@noop {} {\emph {\bibinfo {title} {Quantum Phase Transitions}}},\ \bibinfo {edition} {2nd}\ ed.\ (\bibinfo  {publisher} {Cambridge University Press},\ \bibinfo {address} {Cambridge, U.K.},\ \bibinfo {year} {2011})\BibitemShut {NoStop}%
\bibitem [{\citenamefont {Witczak-Krempa}\ \emph {et~al.}(2014)\citenamefont {Witczak-Krempa}, \citenamefont {S{\o}rensen},\ and\ \citenamefont {Sachdev}}]{witczak2014dynamics}%
  \BibitemOpen
  \bibfield  {author} {\bibinfo {author} {\bibfnamefont {W.}~\bibnamefont {Witczak-Krempa}}, \bibinfo {author} {\bibfnamefont {E.~S.}\ \bibnamefont {S{\o}rensen}},\ and\ \bibinfo {author} {\bibfnamefont {S.}~\bibnamefont {Sachdev}},\ }\href@noop {} {\bibfield  {journal} {\bibinfo  {journal} {Nature Physics}\ }\textbf {\bibinfo {volume} {10}},\ \bibinfo {pages} {361} (\bibinfo {year} {2014})}\BibitemShut {NoStop}%
\bibitem [{\citenamefont {Chen}\ \emph {et~al.}(2014)\citenamefont {Chen}, \citenamefont {Liu}, \citenamefont {Deng}, \citenamefont {Pollet},\ and\ \citenamefont {Prokof‚Äôev}}]{chen2014universal}%
  \BibitemOpen
  \bibfield  {author} {\bibinfo {author} {\bibfnamefont {K.}~\bibnamefont {Chen}}, \bibinfo {author} {\bibfnamefont {L.}~\bibnamefont {Liu}}, \bibinfo {author} {\bibfnamefont {Y.}~\bibnamefont {Deng}}, \bibinfo {author} {\bibfnamefont {L.}~\bibnamefont {Pollet}},\ and\ \bibinfo {author} {\bibfnamefont {N.}~\bibnamefont {Prokof‚Äôev}},\ }\href@noop {} {\bibfield  {journal} {\bibinfo  {journal} {Physical review letters}\ }\textbf {\bibinfo {volume} {112}},\ \bibinfo {pages} {030402} (\bibinfo {year} {2014})}\BibitemShut {NoStop}%
\bibitem [{\citenamefont {Katz}\ \emph {et~al.}(2014)\citenamefont {Katz}, \citenamefont {Sachdev}, \citenamefont {S\o{}rensen},\ and\ \citenamefont {Witczak-Krempa}}]{PhysRevB.90.245109}%
  \BibitemOpen
  \bibfield  {author} {\bibinfo {author} {\bibfnamefont {E.}~\bibnamefont {Katz}}, \bibinfo {author} {\bibfnamefont {S.}~\bibnamefont {Sachdev}}, \bibinfo {author} {\bibfnamefont {E.~S.}\ \bibnamefont {S\o{}rensen}},\ and\ \bibinfo {author} {\bibfnamefont {W.}~\bibnamefont {Witczak-Krempa}},\ }\href {https://doi.org/10.1103/PhysRevB.90.245109} {\bibfield  {journal} {\bibinfo  {journal} {Phys. Rev. B}\ }\textbf {\bibinfo {volume} {90}},\ \bibinfo {pages} {245109} (\bibinfo {year} {2014})}\BibitemShut {NoStop}%
\bibitem [{\citenamefont {Kos}\ \emph {et~al.}(2015)\citenamefont {Kos}, \citenamefont {Poland}, \citenamefont {Simmons-Duffin},\ and\ \citenamefont {Vichi}}]{kos2015bootstrapping}%
  \BibitemOpen
  \bibfield  {author} {\bibinfo {author} {\bibfnamefont {F.}~\bibnamefont {Kos}}, \bibinfo {author} {\bibfnamefont {D.}~\bibnamefont {Poland}}, \bibinfo {author} {\bibfnamefont {D.}~\bibnamefont {Simmons-Duffin}},\ and\ \bibinfo {author} {\bibfnamefont {A.}~\bibnamefont {Vichi}},\ }\href@noop {} {\bibfield  {journal} {\bibinfo  {journal} {Journal of High Energy Physics}\ }\textbf {\bibinfo {volume} {2015}},\ \bibinfo {pages} {106} (\bibinfo {year} {2015})}\BibitemShut {NoStop}%
\bibitem [{\citenamefont {Ran{\c{c}}on}\ and\ \citenamefont {Dupuis}(2014)}]{ranccon2014higgs}%
  \BibitemOpen
  \bibfield  {author} {\bibinfo {author} {\bibfnamefont {A.}~\bibnamefont {Ran{\c{c}}on}}\ and\ \bibinfo {author} {\bibfnamefont {N.}~\bibnamefont {Dupuis}},\ }\href@noop {} {\bibfield  {journal} {\bibinfo  {journal} {Physical Review B}\ }\textbf {\bibinfo {volume} {89}},\ \bibinfo {pages} {180501} (\bibinfo {year} {2014})}\BibitemShut {NoStop}%
\bibitem [{\citenamefont {Rose}\ \emph {et~al.}(2015)\citenamefont {Rose}, \citenamefont {L{\'e}onard},\ and\ \citenamefont {Dupuis}}]{rose2015higgs}%
  \BibitemOpen
  \bibfield  {author} {\bibinfo {author} {\bibfnamefont {F.}~\bibnamefont {Rose}}, \bibinfo {author} {\bibfnamefont {F.}~\bibnamefont {L{\'e}onard}},\ and\ \bibinfo {author} {\bibfnamefont {N.}~\bibnamefont {Dupuis}},\ }\href@noop {} {\bibfield  {journal} {\bibinfo  {journal} {Physical Review B}\ }\textbf {\bibinfo {volume} {91}},\ \bibinfo {pages} {224501} (\bibinfo {year} {2015})}\BibitemShut {NoStop}%
\bibitem [{\citenamefont {Wegner}(1972)}]{wegner1972corrections}%
  \BibitemOpen
  \bibfield  {author} {\bibinfo {author} {\bibfnamefont {F.~J.}\ \bibnamefont {Wegner}},\ }\href@noop {} {\bibfield  {journal} {\bibinfo  {journal} {Physical Review B}\ }\textbf {\bibinfo {volume} {5}},\ \bibinfo {pages} {4529} (\bibinfo {year} {1972})}\BibitemShut {NoStop}%
\bibitem [{\citenamefont {Gazit}\ \emph {et~al.}(2013{\natexlab{a}})\citenamefont {Gazit}, \citenamefont {Podolsky},\ and\ \citenamefont {Auerbach}}]{GPA}%
  \BibitemOpen
  \bibfield  {author} {\bibinfo {author} {\bibfnamefont {S.}~\bibnamefont {Gazit}}, \bibinfo {author} {\bibfnamefont {D.}~\bibnamefont {Podolsky}},\ and\ \bibinfo {author} {\bibfnamefont {A.}~\bibnamefont {Auerbach}},\ }\href {https://doi.org/10.1103/PhysRevLett.110.140401} {\bibfield  {journal} {\bibinfo  {journal} {Phys. Rev. Lett.}\ }\textbf {\bibinfo {volume} {110}},\ \bibinfo {pages} {140401} (\bibinfo {year} {2013}{\natexlab{a}})}\BibitemShut {NoStop}%
\bibitem [{\citenamefont {Gazit}\ \emph {et~al.}(2013{\natexlab{b}})\citenamefont {Gazit}, \citenamefont {Podolsky}, \citenamefont {Auerbach},\ and\ \citenamefont {Arovas}}]{GPAA}%
  \BibitemOpen
  \bibfield  {author} {\bibinfo {author} {\bibfnamefont {S.}~\bibnamefont {Gazit}}, \bibinfo {author} {\bibfnamefont {D.}~\bibnamefont {Podolsky}}, \bibinfo {author} {\bibfnamefont {A.}~\bibnamefont {Auerbach}},\ and\ \bibinfo {author} {\bibfnamefont {D.~P.}\ \bibnamefont {Arovas}},\ }\href {https://doi.org/10.1103/PhysRevB.88.235108} {\bibfield  {journal} {\bibinfo  {journal} {Phys. Rev. B}\ }\textbf {\bibinfo {volume} {88}},\ \bibinfo {pages} {235108} (\bibinfo {year} {2013}{\natexlab{b}})}\BibitemShut {NoStop}%
\bibitem [{\citenamefont {Lucas}\ \emph {et~al.}(2017)\citenamefont {Lucas}, \citenamefont {Gazit}, \citenamefont {Podolsky},\ and\ \citenamefont {Witczak-Krempa}}]{lucas2017dynamical}%
  \BibitemOpen
  \bibfield  {author} {\bibinfo {author} {\bibfnamefont {A.}~\bibnamefont {Lucas}}, \bibinfo {author} {\bibfnamefont {S.}~\bibnamefont {Gazit}}, \bibinfo {author} {\bibfnamefont {D.}~\bibnamefont {Podolsky}},\ and\ \bibinfo {author} {\bibfnamefont {W.}~\bibnamefont {Witczak-Krempa}},\ }\href@noop {} {\bibfield  {journal} {\bibinfo  {journal} {Physical Review Letters}\ }\textbf {\bibinfo {volume} {118}},\ \bibinfo {pages} {056601} (\bibinfo {year} {2017})}\BibitemShut {NoStop}%
\bibitem [{\citenamefont {Symanzik}(1983)}]{SYMANZIK1983187}%
  \BibitemOpen
  \bibfield  {author} {\bibinfo {author} {\bibfnamefont {K.}~\bibnamefont {Symanzik}},\ }\href {https://doi.org/https://doi.org/10.1016/0550-3213(83)90468-6} {\bibfield  {journal} {\bibinfo  {journal} {Nuclear Physics B}\ }\textbf {\bibinfo {volume} {226}},\ \bibinfo {pages} {187 } (\bibinfo {year} {1983})}\BibitemShut {NoStop}%
\bibitem [{\citenamefont {Alford}\ \emph {et~al.}(1995)\citenamefont {Alford}, \citenamefont {Dimm}, \citenamefont {Lepage}, \citenamefont {Hockney},\ and\ \citenamefont {Mackenzie}}]{alford1995lattice}%
  \BibitemOpen
  \bibfield  {author} {\bibinfo {author} {\bibfnamefont {M.}~\bibnamefont {Alford}}, \bibinfo {author} {\bibfnamefont {W.}~\bibnamefont {Dimm}}, \bibinfo {author} {\bibfnamefont {G.}~\bibnamefont {Lepage}}, \bibinfo {author} {\bibfnamefont {G.}~\bibnamefont {Hockney}},\ and\ \bibinfo {author} {\bibfnamefont {P.}~\bibnamefont {Mackenzie}},\ }\href@noop {} {\bibfield  {journal} {\bibinfo  {journal} {Physics Letters B}\ }\textbf {\bibinfo {volume} {361}},\ \bibinfo {pages} {87} (\bibinfo {year} {1995})}\BibitemShut {NoStop}%
\bibitem [{\citenamefont {Hasenbusch}(2001)}]{hasenbusch_eliminating_2001}%
  \BibitemOpen
  \bibfield  {author} {\bibinfo {author} {\bibfnamefont {M.}~\bibnamefont {Hasenbusch}},\ }\href {https://doi.org/10.1088/0305-4470/34/40/302} {\bibfield  {journal} {\bibinfo  {journal} {Journal of Physics A: Mathematical and General}\ }\textbf {\bibinfo {volume} {34}},\ \bibinfo {pages} {8221} (\bibinfo {year} {2001})}\BibitemShut {NoStop}%
\bibitem [{\citenamefont {Prokof'ev}\ and\ \citenamefont {Svistunov}(2001)}]{Prokofev2001WormAlgorithms}%
  \BibitemOpen
  \bibfield  {author} {\bibinfo {author} {\bibfnamefont {N.}~\bibnamefont {Prokof'ev}}\ and\ \bibinfo {author} {\bibfnamefont {B.}~\bibnamefont {Svistunov}},\ }\href {https://doi.org/10.1103/PhysRevLett.87.160601} {\bibfield  {journal} {\bibinfo  {journal} {Phys. Rev. Lett.}\ }\textbf {\bibinfo {volume} {87}},\ \bibinfo {pages} {160601} (\bibinfo {year} {2001})}\BibitemShut {NoStop}%
\bibitem [{\citenamefont {Levin}\ and\ \citenamefont {Nave}(2007)}]{Levin2007TensorRenormalizationGroup}%
  \BibitemOpen
  \bibfield  {author} {\bibinfo {author} {\bibfnamefont {M.}~\bibnamefont {Levin}}\ and\ \bibinfo {author} {\bibfnamefont {C.~P.}\ \bibnamefont {Nave}},\ }\href {https://doi.org/10.1103/PhysRevLett.99.120601} {\bibfield  {journal} {\bibinfo  {journal} {Phys. Rev. Lett.}\ }\textbf {\bibinfo {volume} {99}},\ \bibinfo {pages} {120601} (\bibinfo {year} {2007})}\BibitemShut {NoStop}%
\bibitem [{\citenamefont {Xie}\ \emph {et~al.}(2012)\citenamefont {Xie}, \citenamefont {Chen}, \citenamefont {Qin}, \citenamefont {Zhu}, \citenamefont {Yang},\ and\ \citenamefont {Xiang}}]{Xie2012RGbyHOSVD}%
  \BibitemOpen
  \bibfield  {author} {\bibinfo {author} {\bibfnamefont {Z.~Y.}\ \bibnamefont {Xie}}, \bibinfo {author} {\bibfnamefont {J.}~\bibnamefont {Chen}}, \bibinfo {author} {\bibfnamefont {M.~P.}\ \bibnamefont {Qin}}, \bibinfo {author} {\bibfnamefont {J.~W.}\ \bibnamefont {Zhu}}, \bibinfo {author} {\bibfnamefont {L.~P.}\ \bibnamefont {Yang}},\ and\ \bibinfo {author} {\bibfnamefont {T.}~\bibnamefont {Xiang}},\ }\href {https://doi.org/10.1103/PhysRevB.86.045139} {\bibfield  {journal} {\bibinfo  {journal} {Phys. Rev. B}\ }\textbf {\bibinfo {volume} {86}},\ \bibinfo {pages} {045139} (\bibinfo {year} {2012})}\BibitemShut {NoStop}%
\bibitem [{\citenamefont {Xu}\ \emph {et~al.}(2019)\citenamefont {Xu}, \citenamefont {Sun}, \citenamefont {Lv},\ and\ \citenamefont {Deng}}]{XuWanwan2019HighprecisionMonteCarlostudy}%
  \BibitemOpen
  \bibfield  {author} {\bibinfo {author} {\bibfnamefont {W.}~\bibnamefont {Xu}}, \bibinfo {author} {\bibfnamefont {Y.}~\bibnamefont {Sun}}, \bibinfo {author} {\bibfnamefont {J.-P.}\ \bibnamefont {Lv}},\ and\ \bibinfo {author} {\bibfnamefont {Y.}~\bibnamefont {Deng}},\ }\href {https://doi.org/10.1103/PhysRevB.100.064525} {\bibfield  {journal} {\bibinfo  {journal} {Phys. Rev. B}\ }\textbf {\bibinfo {volume} {100}},\ \bibinfo {pages} {064525} (\bibinfo {year} {2019})}\BibitemShut {NoStop}%
\bibitem [{\citenamefont {Hasenbusch}(2019)}]{Hasenbusch2019MonteCarloClock}%
  \BibitemOpen
  \bibfield  {author} {\bibinfo {author} {\bibfnamefont {M.}~\bibnamefont {Hasenbusch}},\ }\href {https://doi.org/10.1103/PhysRevB.100.224517} {\bibfield  {journal} {\bibinfo  {journal} {Phys. Rev. B}\ }\textbf {\bibinfo {volume} {100}},\ \bibinfo {pages} {224517} (\bibinfo {year} {2019})}\BibitemShut {NoStop}%
\bibitem [{\citenamefont {Ueda}\ \emph {et~al.}(2020)\citenamefont {Ueda}, \citenamefont {Okunishi}, \citenamefont {Harada}, \citenamefont {Kr\ifmmode~\check{c}\else \v{c}\fi{}m\'ar}, \citenamefont {Gendiar}, \citenamefont {Yunoki},\ and\ \citenamefont {Nishino}}]{Ueda2020FiniteScalingAnalysisBKT}%
  \BibitemOpen
  \bibfield  {author} {\bibinfo {author} {\bibfnamefont {H.}~\bibnamefont {Ueda}}, \bibinfo {author} {\bibfnamefont {K.}~\bibnamefont {Okunishi}}, \bibinfo {author} {\bibfnamefont {K.}~\bibnamefont {Harada}}, \bibinfo {author} {\bibfnamefont {R.}~\bibnamefont {Kr\ifmmode~\check{c}\else \v{c}\fi{}m\'ar}}, \bibinfo {author} {\bibfnamefont {A.}~\bibnamefont {Gendiar}}, \bibinfo {author} {\bibfnamefont {S.}~\bibnamefont {Yunoki}},\ and\ \bibinfo {author} {\bibfnamefont {T.}~\bibnamefont {Nishino}},\ }\href {https://doi.org/10.1103/PhysRevE.101.062111} {\bibfield  {journal} {\bibinfo  {journal} {Phys. Rev. E}\ }\textbf {\bibinfo {volume} {101}},\ \bibinfo {pages} {062111} (\bibinfo {year} {2020})}\BibitemShut {NoStop}%
\bibitem [{\citenamefont {Ueda}\ \emph {et~al.}(2017)\citenamefont {Ueda}, \citenamefont {Okunishi}, \citenamefont {Kr\ifmmode~\check{c}\else \v{c}\fi{}m\'ar}, \citenamefont {Gendiar}, \citenamefont {Yunoki},\ and\ \citenamefont {Nishino}}]{Ueda2017CriticalBehaviorIcosahedron}%
  \BibitemOpen
  \bibfield  {author} {\bibinfo {author} {\bibfnamefont {H.}~\bibnamefont {Ueda}}, \bibinfo {author} {\bibfnamefont {K.}~\bibnamefont {Okunishi}}, \bibinfo {author} {\bibfnamefont {R.}~\bibnamefont {Kr\ifmmode~\check{c}\else \v{c}\fi{}m\'ar}}, \bibinfo {author} {\bibfnamefont {A.}~\bibnamefont {Gendiar}}, \bibinfo {author} {\bibfnamefont {S.}~\bibnamefont {Yunoki}},\ and\ \bibinfo {author} {\bibfnamefont {T.}~\bibnamefont {Nishino}},\ }\href {https://doi.org/10.1103/PhysRevE.96.062112} {\bibfield  {journal} {\bibinfo  {journal} {Phys. Rev. E}\ }\textbf {\bibinfo {volume} {96}},\ \bibinfo {pages} {062112} (\bibinfo {year} {2017})}\BibitemShut {NoStop}%
\bibitem [{\citenamefont {Ueda}\ \emph {et~al.}(2014)\citenamefont {Ueda}, \citenamefont {Okunishi},\ and\ \citenamefont {Nishino}}]{Ueda2014DoublingEntanglementSpectrum}%
  \BibitemOpen
  \bibfield  {author} {\bibinfo {author} {\bibfnamefont {H.}~\bibnamefont {Ueda}}, \bibinfo {author} {\bibfnamefont {K.}~\bibnamefont {Okunishi}},\ and\ \bibinfo {author} {\bibfnamefont {T.}~\bibnamefont {Nishino}},\ }\href {https://doi.org/10.1103/PhysRevB.89.075116} {\bibfield  {journal} {\bibinfo  {journal} {Phys. Rev. B}\ }\textbf {\bibinfo {volume} {89}},\ \bibinfo {pages} {075116} (\bibinfo {year} {2014})}\BibitemShut {NoStop}%
\bibitem [{\citenamefont {Ueda}\ and\ \citenamefont {Oshikawa}(2023)}]{Ueda2023Finite-sizeBondDimensionEffectsTNR}%
  \BibitemOpen
  \bibfield  {author} {\bibinfo {author} {\bibfnamefont {A.}~\bibnamefont {Ueda}}\ and\ \bibinfo {author} {\bibfnamefont {M.}~\bibnamefont {Oshikawa}},\ }\href {https://doi.org/10.1103/PhysRevB.108.024413} {\bibfield  {journal} {\bibinfo  {journal} {Phys. Rev. B}\ }\textbf {\bibinfo {volume} {108}},\ \bibinfo {pages} {024413} (\bibinfo {year} {2023})}\BibitemShut {NoStop}%
\bibitem [{\citenamefont {Tagliacozzo}\ \emph {et~al.}(2008)\citenamefont {Tagliacozzo}, \citenamefont {de~Oliveira}, \citenamefont {Iblisdir},\ and\ \citenamefont {Latorre}}]{Tagliacozzo2008ScalingEntanglementMPS}%
  \BibitemOpen
  \bibfield  {author} {\bibinfo {author} {\bibfnamefont {L.}~\bibnamefont {Tagliacozzo}}, \bibinfo {author} {\bibfnamefont {T.~R.}\ \bibnamefont {de~Oliveira}}, \bibinfo {author} {\bibfnamefont {S.}~\bibnamefont {Iblisdir}},\ and\ \bibinfo {author} {\bibfnamefont {J.~I.}\ \bibnamefont {Latorre}},\ }\href {https://doi.org/10.1103/PhysRevB.78.024410} {\bibfield  {journal} {\bibinfo  {journal} {Phys. Rev. B}\ }\textbf {\bibinfo {volume} {78}},\ \bibinfo {pages} {024410} (\bibinfo {year} {2008})}\BibitemShut {NoStop}%
\bibitem [{\citenamefont {Pollmann}\ \emph {et~al.}(2009)\citenamefont {Pollmann}, \citenamefont {Mukerjee}, \citenamefont {Turner},\ and\ \citenamefont {Moore}}]{Pollmann2009TheoryFiniteEntanglementScaling}%
  \BibitemOpen
  \bibfield  {author} {\bibinfo {author} {\bibfnamefont {F.}~\bibnamefont {Pollmann}}, \bibinfo {author} {\bibfnamefont {S.}~\bibnamefont {Mukerjee}}, \bibinfo {author} {\bibfnamefont {A.~M.}\ \bibnamefont {Turner}},\ and\ \bibinfo {author} {\bibfnamefont {J.~E.}\ \bibnamefont {Moore}},\ }\href {https://doi.org/10.1103/PhysRevLett.102.255701} {\bibfield  {journal} {\bibinfo  {journal} {Phys. Rev. Lett.}\ }\textbf {\bibinfo {volume} {102}},\ \bibinfo {pages} {255701} (\bibinfo {year} {2009})}\BibitemShut {NoStop}%
\bibitem [{\citenamefont {Calabrese}\ and\ \citenamefont {Lefevre}(2008)}]{Calabrese2008Entanglementspectrum}%
  \BibitemOpen
  \bibfield  {author} {\bibinfo {author} {\bibfnamefont {P.}~\bibnamefont {Calabrese}}\ and\ \bibinfo {author} {\bibfnamefont {A.}~\bibnamefont {Lefevre}},\ }\href {https://doi.org/10.1103/PhysRevA.78.032329} {\bibfield  {journal} {\bibinfo  {journal} {Phys. Rev. A}\ }\textbf {\bibinfo {volume} {78}},\ \bibinfo {pages} {032329} (\bibinfo {year} {2008})}\BibitemShut {NoStop}%
\bibitem [{\citenamefont {Pirvu}\ \emph {et~al.}(2012)\citenamefont {Pirvu}, \citenamefont {Vidal}, \citenamefont {Verstraete},\ and\ \citenamefont {Tagliacozzo}}]{Pirvu2012MatrixProductStateFiniteSize}%
  \BibitemOpen
  \bibfield  {author} {\bibinfo {author} {\bibfnamefont {B.}~\bibnamefont {Pirvu}}, \bibinfo {author} {\bibfnamefont {G.}~\bibnamefont {Vidal}}, \bibinfo {author} {\bibfnamefont {F.}~\bibnamefont {Verstraete}},\ and\ \bibinfo {author} {\bibfnamefont {L.}~\bibnamefont {Tagliacozzo}},\ }\href {https://doi.org/10.1103/PhysRevB.86.075117} {\bibfield  {journal} {\bibinfo  {journal} {Phys. Rev. B}\ }\textbf {\bibinfo {volume} {86}},\ \bibinfo {pages} {075117} (\bibinfo {year} {2012})}\BibitemShut {NoStop}%
\bibitem [{\citenamefont {De~Lathauwer}\ \emph {et~al.}(2000)\citenamefont {De~Lathauwer}, \citenamefont {De~Moor},\ and\ \citenamefont {Vandewalle}}]{DeLathauwer2000HOSVD}%
  \BibitemOpen
  \bibfield  {author} {\bibinfo {author} {\bibfnamefont {L.}~\bibnamefont {De~Lathauwer}}, \bibinfo {author} {\bibfnamefont {B.}~\bibnamefont {De~Moor}},\ and\ \bibinfo {author} {\bibfnamefont {J.}~\bibnamefont {Vandewalle}},\ }\href {https://doi.org/10.1137/S0895479896305696} {\bibfield  {journal} {\bibinfo  {journal} {SIAM Journal on Matrix Analysis and Applications}\ }\textbf {\bibinfo {volume} {21}},\ \bibinfo {pages} {1253} (\bibinfo {year} {2000})},\ \Eprint {https://arxiv.org/abs/https://doi.org/10.1137/S0895479896305696} {https://doi.org/10.1137/S0895479896305696} \BibitemShut {NoStop}%
\bibitem [{\citenamefont {Adachi}\ \emph {et~al.}(2020)\citenamefont {Adachi}, \citenamefont {Okubo},\ and\ \citenamefont {Todo}}]{adachi2020ATRG}%
  \BibitemOpen
  \bibfield  {author} {\bibinfo {author} {\bibfnamefont {D.}~\bibnamefont {Adachi}}, \bibinfo {author} {\bibfnamefont {T.}~\bibnamefont {Okubo}},\ and\ \bibinfo {author} {\bibfnamefont {S.}~\bibnamefont {Todo}},\ }\href {https://doi.org/10.1103/PhysRevB.102.054432} {\bibfield  {journal} {\bibinfo  {journal} {Phys. Rev. B}\ }\textbf {\bibinfo {volume} {102}},\ \bibinfo {pages} {054432} (\bibinfo {year} {2020})}\BibitemShut {NoStop}%
\bibitem [{\citenamefont {Kadoh}\ and\ \citenamefont {Nakayama}(2019)}]{kadoh2019TriagTRG}%
  \BibitemOpen
  \bibfield  {author} {\bibinfo {author} {\bibfnamefont {D.}~\bibnamefont {Kadoh}}\ and\ \bibinfo {author} {\bibfnamefont {K.}~\bibnamefont {Nakayama}},\ }\href@noop {} {\bibinfo {title} {Renormalization group on a triad network}} (\bibinfo {year} {2019}),\ \Eprint {https://arxiv.org/abs/1912.02414} {arXiv:1912.02414 [hep-lat]} \BibitemShut {NoStop}%
\bibitem [{\citenamefont {Bloch}\ \emph {et~al.}(2021)\citenamefont {Bloch}, \citenamefont {Jha}, \citenamefont {Lohmayer},\ and\ \citenamefont {Meister}}]{bloch2021TRGO2model}%
  \BibitemOpen
  \bibfield  {author} {\bibinfo {author} {\bibfnamefont {J.}~\bibnamefont {Bloch}}, \bibinfo {author} {\bibfnamefont {R.~G.}\ \bibnamefont {Jha}}, \bibinfo {author} {\bibfnamefont {R.}~\bibnamefont {Lohmayer}},\ and\ \bibinfo {author} {\bibfnamefont {M.}~\bibnamefont {Meister}},\ }\href {https://doi.org/10.1103/PhysRevD.104.094517} {\bibfield  {journal} {\bibinfo  {journal} {Phys. Rev. D}\ }\textbf {\bibinfo {volume} {104}},\ \bibinfo {pages} {094517} (\bibinfo {year} {2021})}\BibitemShut {NoStop}%
\end{thebibliography}%

\end{document}